\documentclass[aoas, preprint]{imsart}

\usepackage{amsthm,amsmath,natbib}
\usepackage{graphicx}
\RequirePackage{natbib}
\RequirePackage[colorlinks,citecolor=blue,urlcolor=blue]{hyperref}

\usepackage{float}
\usepackage[caption = false]{subfig}

\arxiv{arXiv:0000.0000}

\startlocaldefs
\endlocaldefs

\begin{document}

\begin{frontmatter}

\title{What we look at in paintings: A comparison between experienced and inexperienced art viewers}

\runtitle{Stochastic modeling of eye movements}

\begin{aug}
\author{\fnms{Anna-Kaisa} \snm{Ylitalo}\thanksref{t1}\ead[label=e1]{anna-kaisa.ylitalo@jyu.fi}},
\author{\fnms{Aila} \snm{S\"arkk\"a}\thanksref{t2}\ead[label=e2]{aila@chalmers.se}}
\and
\author{\fnms{Peter} \snm{Guttorp}\thanksref{t3}\ead[label=e3]{peter@stat.washington.edu}}

\thankstext{t1}{Supported in part by the Finnish Doctoral Programme in
Stochastics and Statistics and  by the Academy of Finland (Project number 275929).}
\thankstext{t2}{Supported by the Knut and Alice Wallenberg Foundation.}
\runauthor{A.-K. Ylitalo et al.}

\affiliation{University of Jyvaskyla\thanksmark{t1}, Chalmers
University of Technology and University of
Gothenburg\thanksmark{t2}, University of Washington and Norwegian
Computing Center
\thanksmark{t3} }

\address{A.-K. Ylitalo \\
Department of Music\\
P.O.Box 35\\
FI-40014 University of Jyvaskyla \\
Finland \\
and \\
Department of Mathematics and Statistics \\
P.O.Box 35 (MaD) \\
FI-40014 University of Jyvaskyla \\
Finland \\
\printead{e1}\\
\phantom{E-mail:\ } 
}

\address{A. S\"arkk\"a \\
Mathematical Sciences\\
Chalmers University of Technology \\ \, and University of Gothenburg\\
SE-412 96 Gothenburg \\ 
Sweden \\
\printead{e2}\\
\phantom{E-mail:\ }}

\address{P. Guttorp \\
Department of Statistics\\
Box 354322\\
University of Washingtion \\
Seattle, WA 98195-4322 \\
USA \\
\printead{e3}\\
\phantom{E-mail:\ }}

\end{aug}

\begin{abstract}
How do people look at art? Are there any differences between how
experienced and inexperienced art viewers look at a painting? We approach these
questions by analyzing and modeling eye movement data from a
cognitive art research experiment, where the eye movements of
twenty test subjects, ten experienced and ten inexperienced art viewers, were
recorded while they were looking at paintings.

Eye movements consist of stops of the gaze as well as jumps
between the stops. Hence, the observed gaze stop locations can be
thought as a spatial point pattern, which can be modeled by a
spatio-temporal point process. We introduce some statistical tools
to analyze the spatio-temporal eye movement data, and compare the
eye movements of experienced and inexperienced art viewers. In addition, we
develop a stochastic model, which is rather simple but fits quite
well to the eye movement data, to further investigate the
differences between the two groups through functional summary statistics.
\end{abstract}

\begin{keyword}
\kwd{Coverage} \kwd{Intensity} \kwd{Point process} \kwd{Shift
function} \kwd{Transition probability}
\end{keyword}

\end{frontmatter}

\section{Introduction}

Eye movements are outcomes of cognitive processes in the human
brain, and can be recorded with high spatial and temporal
resolution by computerized eye trackers. Eye movements provide
valuable information about cognitive processes
\citep*{duchowski,rayner,rayner_2009}, and tracking of them is
analyzed in a range of different areas, such as language and music
reading \citep*{rayner,kinsler}, psychology \citep*{Findlay}, and
marketing research \citep*{nagasawa}.

The first measurements of eye movements were made in 1879
independently by M. Lamare in France and Ewald Hering in Germany
\citep*{wade2010}. Both researchers used an acoustic approach, and
noticed, much to everyone's surprise, that in reading text the
gaze moves in jerks between points of rest. These extremely rapid
movements are called {\em saccades}. They are essentially
involuntary, so once a saccade starts it cannot be interrupted
\citep{Findlay}. Thus researchers can predict saccade lengths from
very few observations early in the saccade \citep*{komogertsev}.
The points of rest, periods in which the gaze is staying
relatively still around a location of the target space, are called
{\em fixations} \citep*{barlow}. Eye movements can thus be
represented as an alternating sequence of fixations and saccades.
In this paper we will mainly focus on sequences of fixation
locations and times in the target space, and call such a sequence
of observations a fixation process. Our targets are pictures of
paintings and the data consist of recorded eye movements of
subjects on the paintings. We will introduce some new statistical
tools and a model to analyze the fixation process by means of group comparisons.

The fixation process is regarded as a spatio-temporal point
process. We are aware of only few studies that use point process
methodology to analyze fixation processes. \cite*{barthelme} use
inhomogeneous Poisson processes to model fixation locations, and
\cite*{engbert} perform some preliminary analysis of clustering of
the fixations by using inhomogeneous pair correlation function.
Furthermore, they construct a dynamic model for saccade selection
by discretizing the space.

We are interested in studying how people look at art. The first
such study was carried out using two movie cameras by
\cite*{buswell}. His findings indicated that much of the theory of
how people look at art needed reformulation. Another interesting
eye movement study related to arts is brought out by
\cite*{locher}. He developed a two-stage model for art viewing,
where the first stage of viewing of a painting was to obtain a
general view of its structure and semantics. In Locher's study,
subjects describing a painting orally while viewing it started to
describe the painting only after a couple of seconds after they
had begun to look at it. At the second stage of the model the
viewer focuses on some interesting features, which are analysed
from an aesthetic point of view.

In this paper, we concentrate on comparing the fixation process of inexperienced and experienced art viewers, {\em novices} and {\em non-novices} for short. Already in his pioneering work
\cite{buswell} compared the fixation durations of 61 non-novices and 117 novices (originally "art experts" and "lay persons") by comparing the group averages based on the 25
first fixations. He repeated the experiment for seven paintings,
and in each case he concluded that the non-novices made
shorter fixations than the novices. Nevertheless, when comparing the
fixations in the areas of subdivided picture, he did not report
any major differences between the two groups. However,
there are some studies where some differences have been found. In
\cite*{kristjanson} ten paintings were shown to a group of non-novices and a group of novices ("artists'' and "non-artists'' in the original paper). All the
subjects were familiar with three of the paintings (same for all
subjects), the non-novices group was familiar with another three
paintings while the remaining paintings were unknown to all
subjects. The only statistically significant difference between
the non-novice and novice groups found was that non-novices made more
fixations on the parts of the paintings that were not centers of
interest when viewing unfamiliar paintings compared to novices. The authors concluded that the non-novices made a more
thorough investigation of all areas of the unfamiliar paintings
than the novices. The final conclusion was that it is
conceivable that the non-novices extract a different amount or
different kind of information during each fixation, even though
the pattern of fixations may vary little from the pattern of novices.

\cite*{vogt} used eye movement patterns to investigate how non-novices and novices ("artists'' and "artistically untrained people" in the original paper) view art. Participants were
shown realistic and abstract works of art under two conditions:
one asking them to free scan the paintings, and the other asking
them to memorize them. Participants' eye movements were tracked as
they either looked at the images or tried to memorize them, and
their recall for the memorized images was recorded. The
researchers found no differences in the fixation frequency or
duration between picture types for non-novices and novices.
However, across the two conditions, the novices had more short
fixations while free scanning the works, and fewer long fixations
while trying to memorize. Non-novices followed the opposite pattern.
In addition, non-novices spend more time than novices by scanning the areas which were
not defined as regions of interest. There was no statistically
significant difference in the recall of the images across groups,
except that non-novices recalled abstract images better than novices, and remembered more pictorial details.

Here, we investigate whether we can find any differences between
how non-novices and novices view a particular painting by
using a new set of tools to study eye movements. As mentioned
earlier, the fixation process is described by a spatio-temporal
point process, and intensities of the point processes and
distributions of the fixation and saccade durations are used to
compare the eye movements of novices and non-novices. We are
mainly concentrating on the eye movements of 20 subjects, 10
novices and 10 non-novices, on the painting {\em Koli
landscape} by Eero J\"arnefelt \citep*{sundell} but give some results based
on the other five paintings as well. For example, we
investigate whether the fixation duration distributions of the novice and non-novice 
groups remain the same from painting to painting. Finally, based on the data analysis we construct
a rather simple stochastic model that further helps as to describe
the fixation process and to investigate differences between the non-novices and novices in terms of functional summary statistics.

The paper is organized as follows. The experimental set-up and the
data from the cognitive art research experiment as well as the
results from the preliminary data analysis are described in
Section \ref{sec:data}. In Section \ref{sec:comparison}, we show
the results of a comparison of the two groups, novices and non-novices. A stochastic model is constructed in Section
\ref{sec:model}, and it is used to make further comparisons of the groups for the J\"arnefelt painting in Section
\ref{sec:results}. Finally, in Section \ref{sec:disc} we discuss
the results.

\section{Data and data analysis}
\label{sec:data}
\subsection{Art experiment}
\begin{figure}
\centering
\begin{tabular}{cc}
\subfloat[Eero J\"arnefelt - Koli-landscape (the turn of 19$^{\text{th}}$ and 20$^{\text{th}}$ century), source: \cite{sundell}]{\includegraphics[width = 0.45\textwidth]{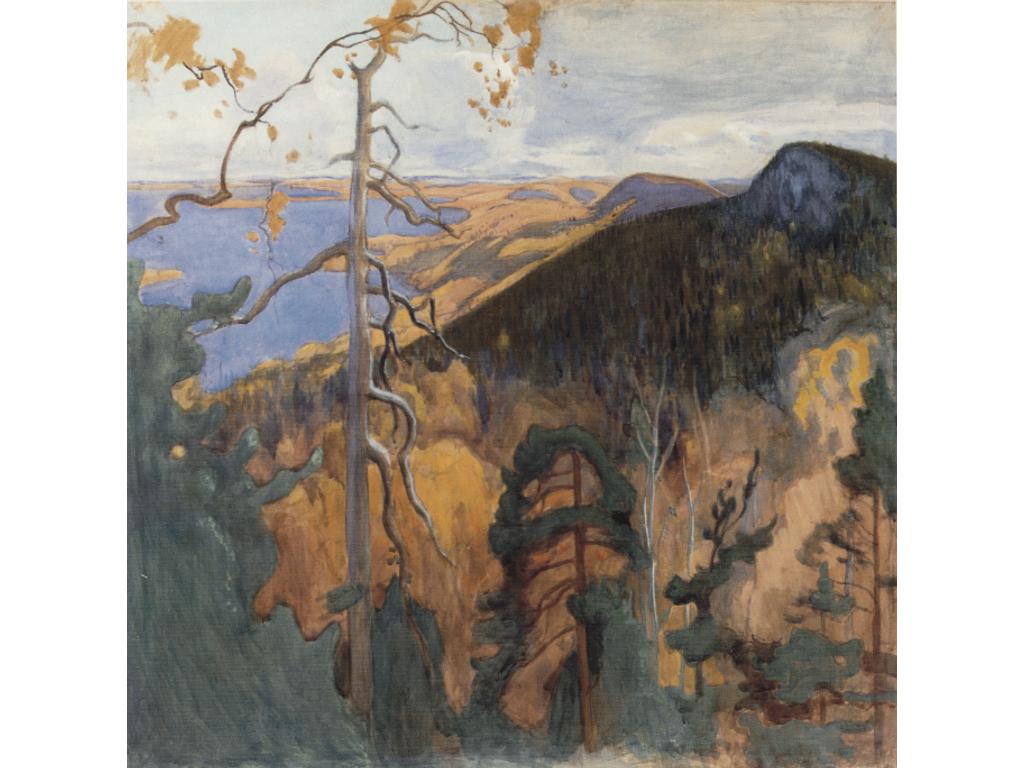}} &
\subfloat[Claude Monet - Terrace at Sainte-Adresse (1867), source: \cite{tuckey}]{\includegraphics[width = 0.45\textwidth]{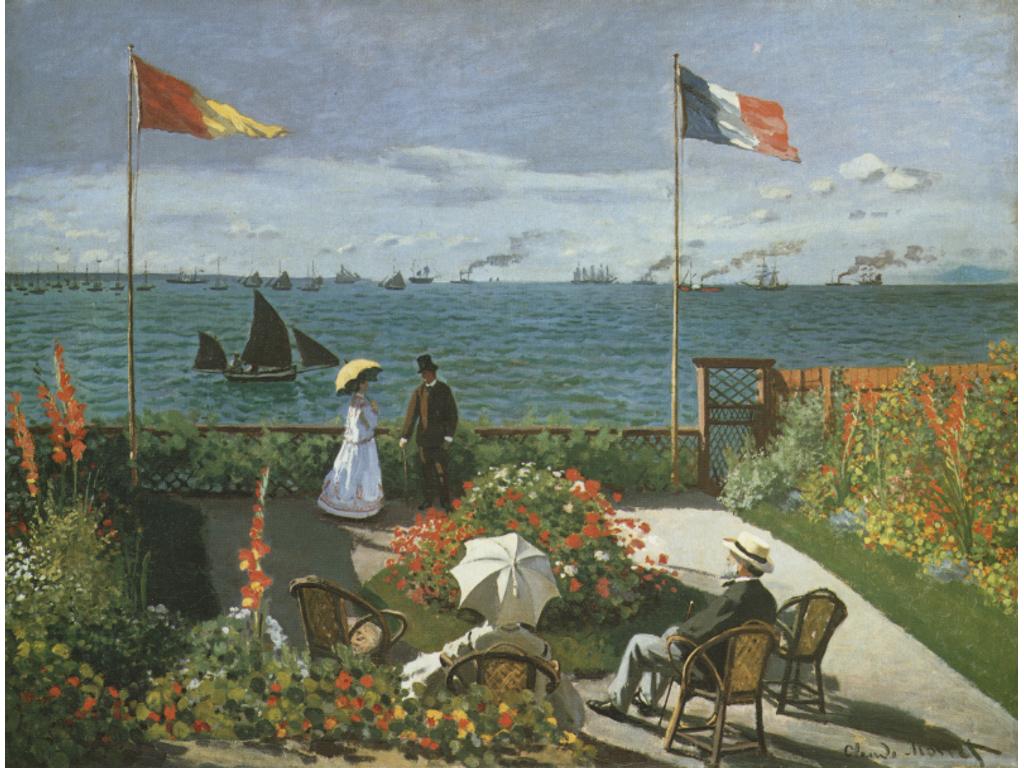}} \\
\subfloat[Risto Suomi - The Cross of Destiny (1988), source: \cite{mikkola}]{\includegraphics[width = 0.45\textwidth]{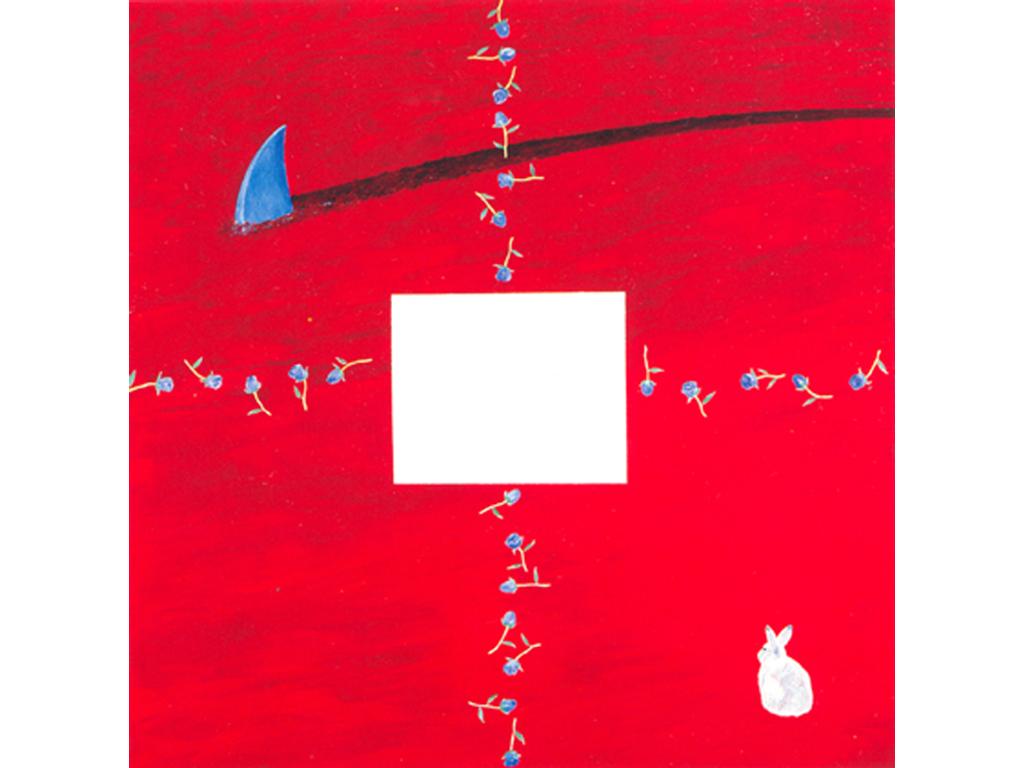}} &
\subfloat[Wassily Kandinsky - Black Bow (1912), source: \cite{duchting}]{\includegraphics[width = 0.45\textwidth]{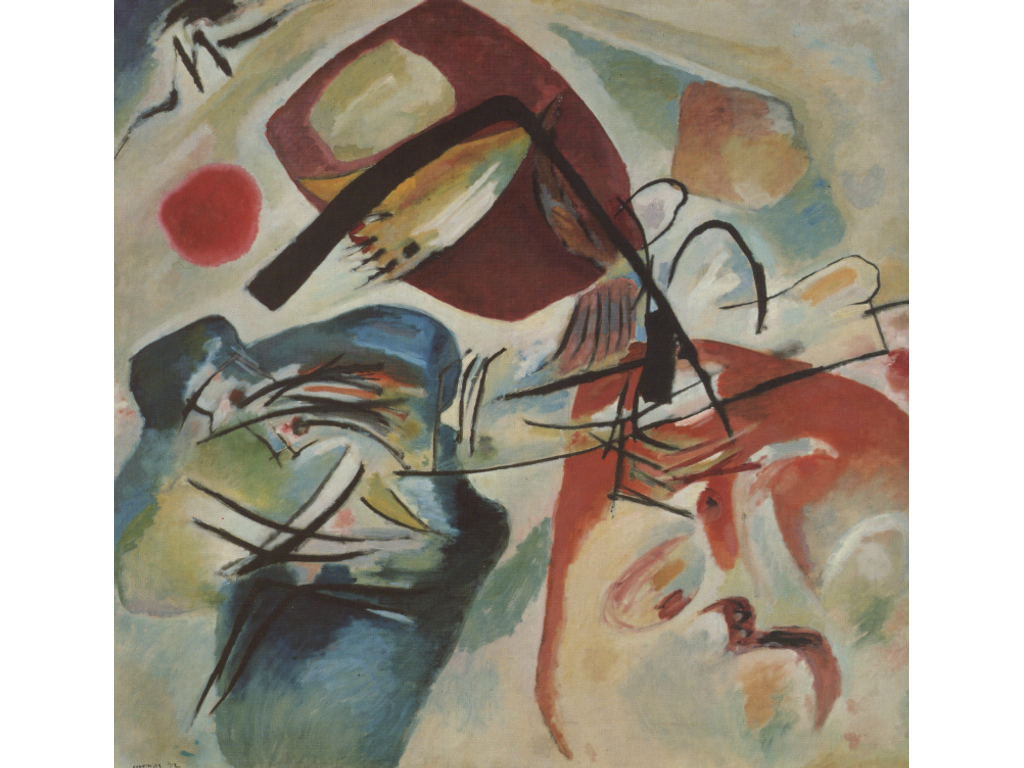}} \\   
\subfloat[Nicolas Poussin - Lamentation over the Death Christ (17$^{\text{th}}$ century), source: \cite{merot}]{\includegraphics[width = 0.45\textwidth]{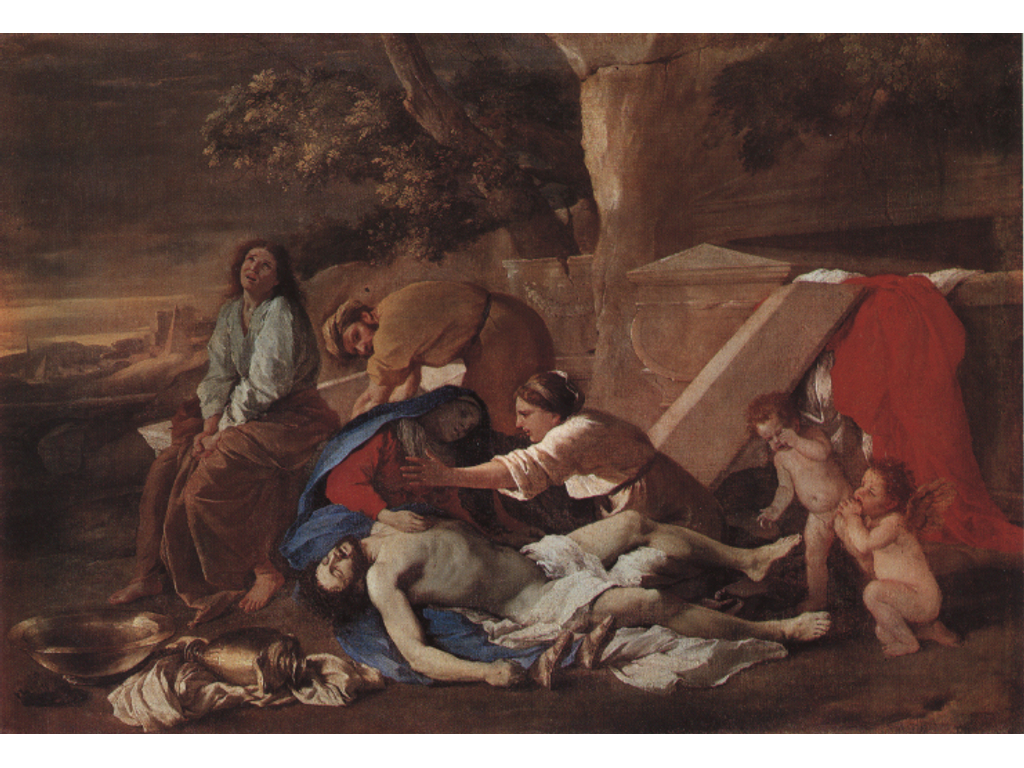}}  & 
\subfloat[Pasi Tammi - Poem Forces to Kneel Down (1999), source: \cite{tammi}]{\includegraphics[width = 0.45\textwidth]{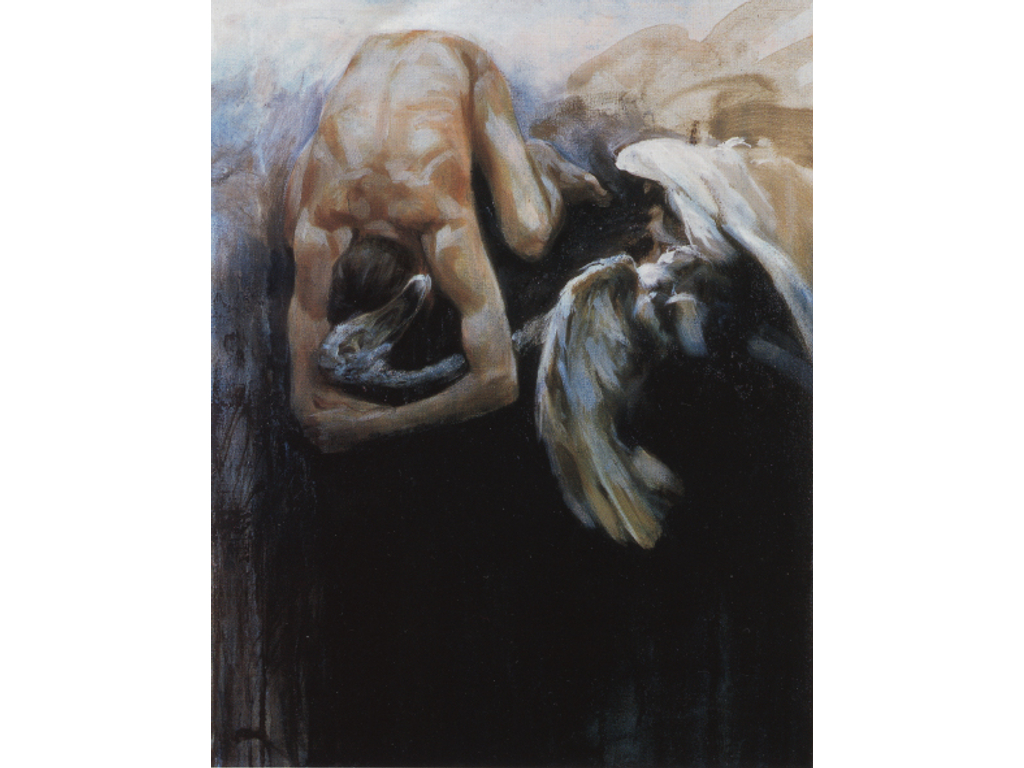}}
\end{tabular}
\caption{The stimulus paintings used in the experiment.}
\label{fig:paintings}
\end{figure}

Twenty test subjects participated in an experiment, where the task
was to observe six pictures of paintings on a computer screen,
each during three minutes, and describe orally the mood in each
painting while their eye movements and voice were recorded.  The paintings can be seen in Figure \ref{fig:paintings}. The
eye movements were recorded by the SMI iView X\texttrademark
Hi-Speed eye tracker with a temporal resolution of 500Hz. The
resolution of the target screen was 1024 $\times$ 768 pixels and
the distance between a participant's head and the screen was
around 85 cm. A forehead rest was used in order to avoid redundant
movement of the head. The data were collected by Pertti Saariluoma
and Sari Kuuva (University of Jyv\"askyl\"a) and Mar\'ia \'Alvarez
Gil (University of Salamanca) with technical help from Jarkko
Hautala and Tuomo Kujala (University of Jyv\"askyl\"a). All
subjects were students at the University of Jyv\"askyl\"a at the
time of the study. Ten of the subjects were either art students
(8) or students who had studied art history and frequently visited
art exhibitions (2). The remaining ten subjects were students who
did not have art as their major nor their hobby. We call the
participants in the first group {\em non-novices} and participants
in the latter group {\em novices}. Five of the participants were
men (three novices and two non-novices) and 15 women (seven
novices and eight non-novices).

\begin{figure}
\centering
  \includegraphics[width=\textwidth]{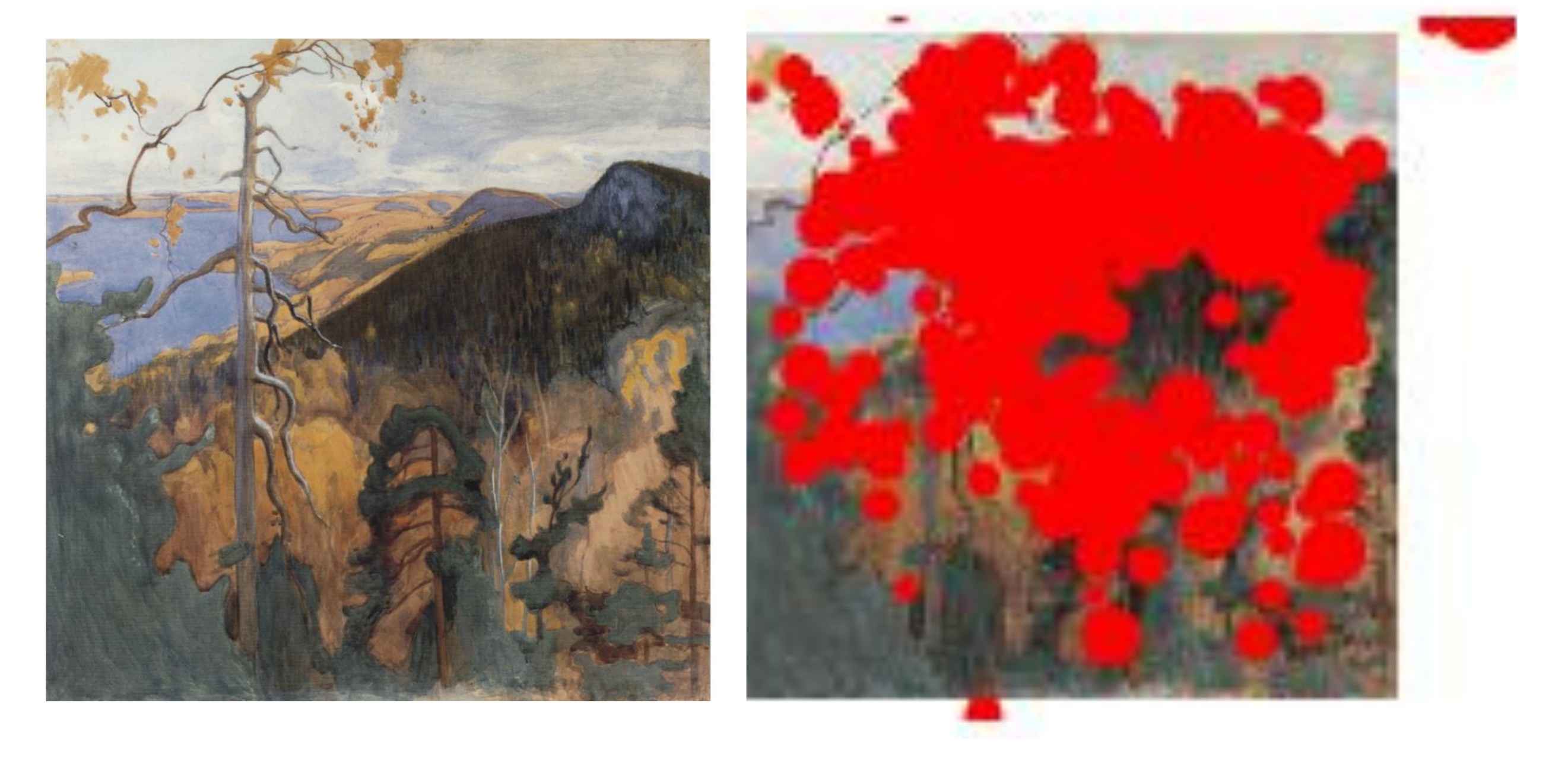}\\
  \caption{(Left image) Koli landscape by Eero J\"arnefelt  from the turn of 19$^{\text{th}}$ and 20$^{\text{th}}$ century. (Right image) Fixations (red {spots}) for Subject 15, a male non-novice. The duration of each fixation is proportional to the size of the {spot}.}
 \label{fig:Koli}
\end{figure}

In this article our main purpose is to introduce some new tools
for eye movement analysis. Therefore, instead of analyzing all the
six paintings in detail we concentrate mainly on the eye movements
on one of the paintings, namely the J\"arnefelt painting Koli
landscape, shown in Figure \ref{fig:paintings}(a). The
resolution of the image of the painting is 770 $\times$ 768
pixels, hence it does not fill the whole computer screen and there
are white areas on both sides of the painting. For some subjects,
some of the fixations were located in these white areas outside
the painting (Figure \ref{fig:Koli}, right) and excluded from the analysis. We will treat saccades going
outside the image as missing values.

\subsection{Data description}

\label{sec:datadescription} 
\begin{figure}
\centering
  \includegraphics[scale=0.35]{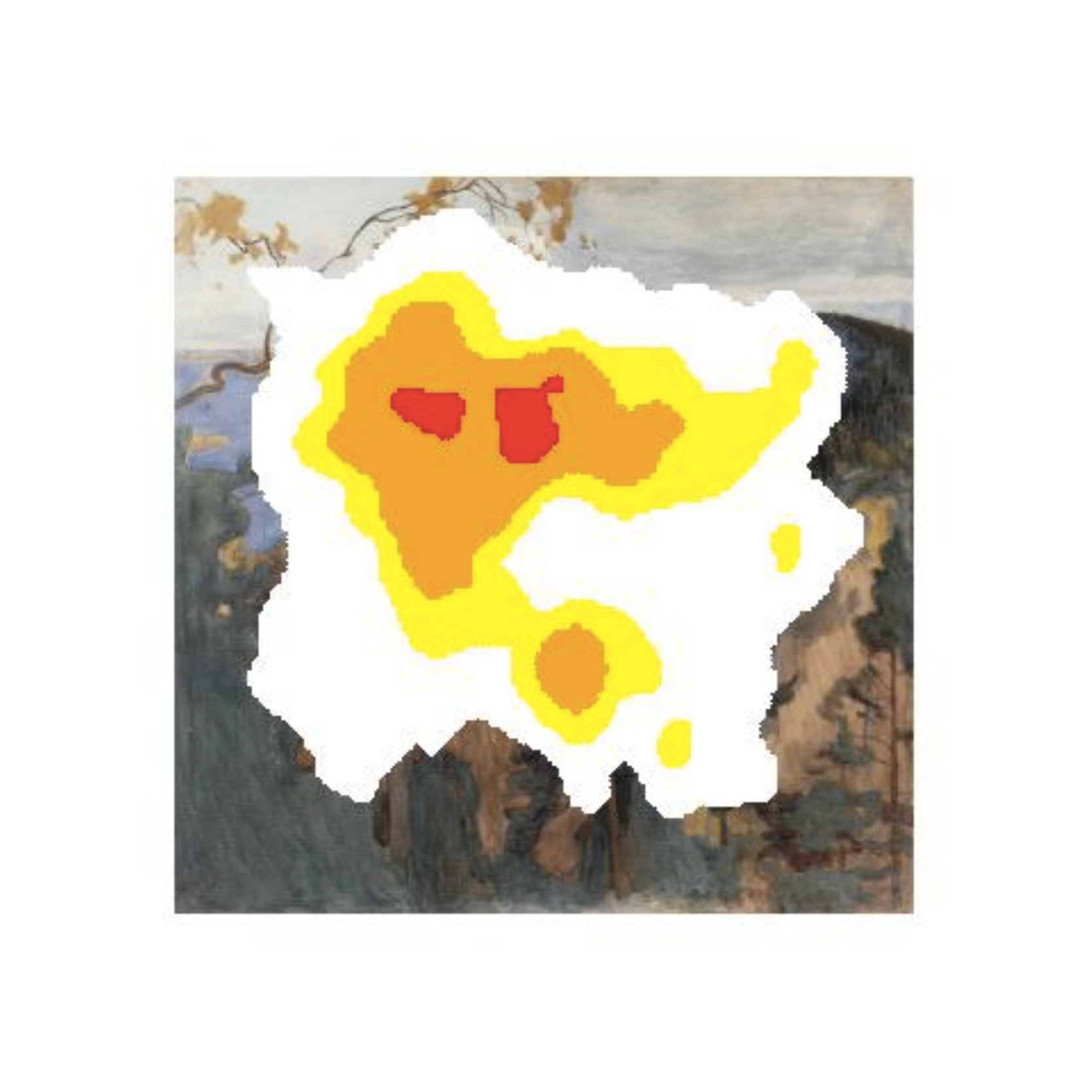}
\includegraphics[scale=0.35]{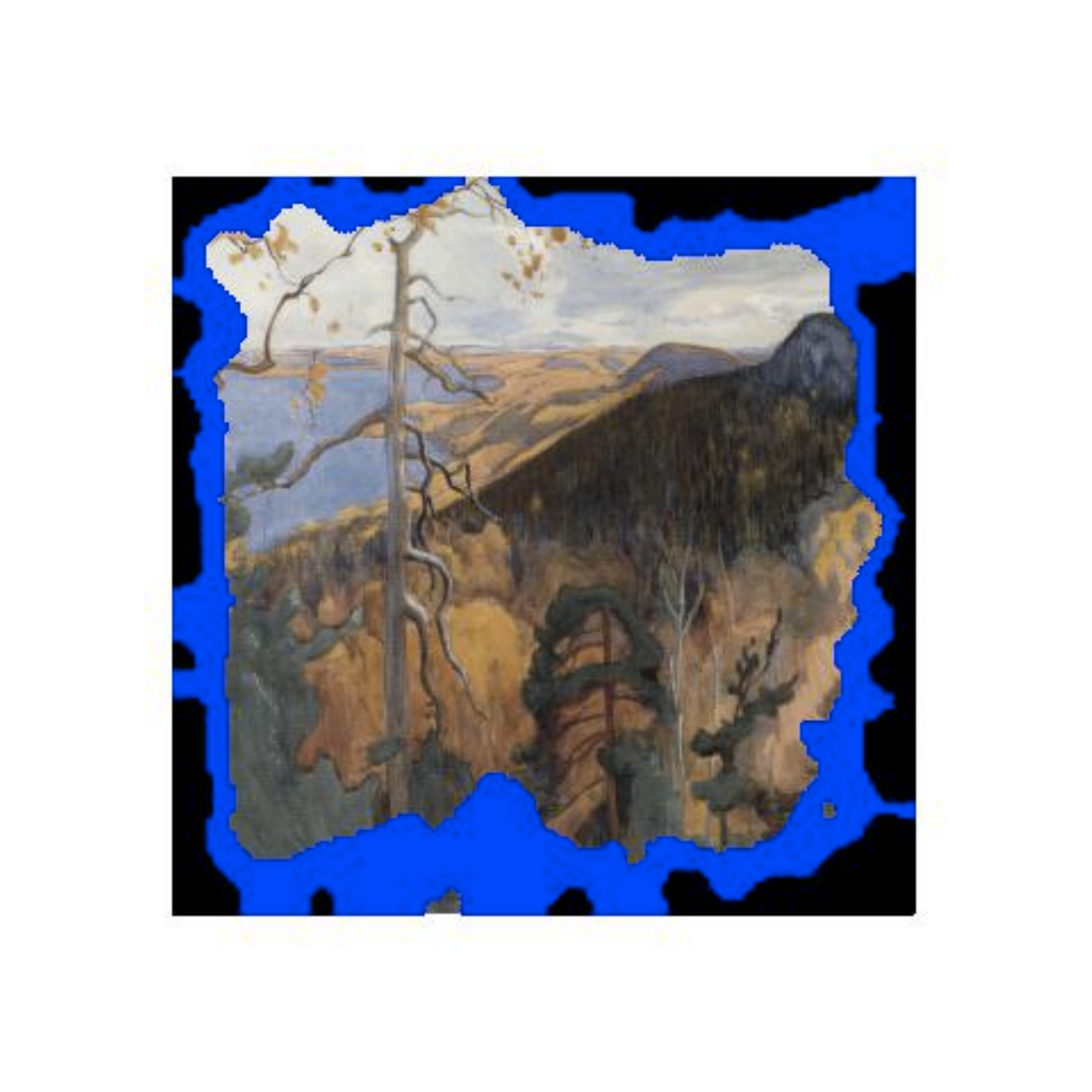}\\
  \caption{Areas of the J\"arnefelt painting that were visited most (left) and least (right) by the gaze of the subjects. (Left image) Top 50\% white, Top 20\% yellow, Top 10\% orange and Top 1\% red. (Right image) Bottom 30\% blue and Bottom 10\% black.}
 \label{fig:KoliIntensityAll}
\end{figure}

We first look at the eye movement data
as a whole and later (Section \ref{sec:comparison}) compare the
eye movements of novices and non-novices. An important way of
describing the viewing of the painting is to look at a smoothed
plot of all fixation locations of all subjects, which represents
the viewing foci of the painting. Technically, when describing the
locations of fixations by a spatial point process this is an
estimate of the intensity of the fixation process, marginalized
over time. We use the R package \verb+spatstat+ for the intensity estimation \citep{baddeley}. A chi-square test for quadrat counts clearly
rejects the hypothesis of constant intensity ($p < 0.001$).
The areas of the painting that have been visited most
by the subjects are shown in Figure \ref{fig:KoliIntensityAll}
(left). We can see that the trunk of the main tree going down
almost the entire length of the painting and the top of the small
tree next to it seem to be the areas of particular interest in the
painting. In Figure \ref{fig:KoliIntensityAll} (right) one can see
the least visited areas, the edges of the painting in this case.

The fixation process varies a lot between subjects, and hence it
is interesting to look at some summary statistics for the data. The supplemental article
\cite*{suppA} includes a table, which shows the total number of
fixations and the number of short fixations (less than 40 ms)
inside the picture with means, medians and standard deviations of
fixation durations for each subject. The number of fixations
during the three minutes time period varies between 326 and 770,
and the median fixation duration varies between 150 ms and 438 ms.
The mean fixation duration is always greater than the median,
indicating a right skewed distribution of the fixation duration.

\begin{figure}
\centering
  \includegraphics[width=\textwidth]{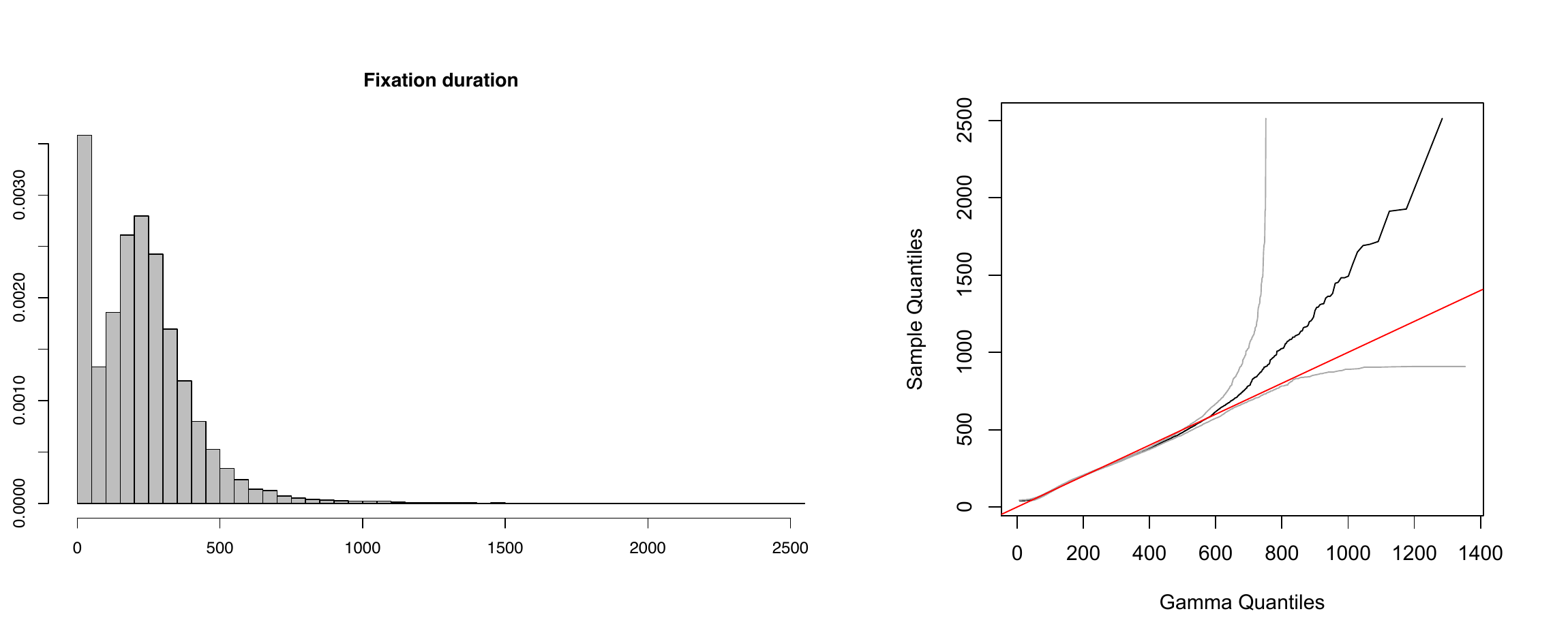}\\
  \caption{(Left panel) Distribution of fixation durations for all subjects, when short fixations are included. (Right panel) Gamma quantile-quantile plot for fixation durations exceeding 40ms with asymptotic 95\% confidence region (grey) and the line (red) of unit slope through the origin corresponding to perfect agreement of empirical and theoretical quantiles. }
  \label{fig:Histogram_fix_dur}
\end{figure}

Short fixations are not included in our analysis because they are
believed to be microsaccades, or eye movements within a fixation
\citep*{manor}. The histogram on the left side of Figure
\ref{fig:Histogram_fix_dur} indicates that the duration
distribution may be a mixture of very short durations and regular
ones. When excluding the short fixations, the remaining durations
seem to follow a gamma distribution. The right panel shows a gamma
quantile-quantile plot \citep*{Gnanadesikan} of the regular
(longer than 40 ms) fixation durations. The red line of slope 1
falls inside the simultaneous confidence band
\citep*{doksumsievers}, indicating that the durations can be
described well by a gamma distribution.

As mentioned earlier, the complete eye movement process consists
of fixations and saccades, the latter being the rapid movements
between the fixations. The supplemental article \cite*{suppB} shows summary statistics for
saccade durations and saccade lengths, i.e., the distances between
consecutive fixation locations. The distribution of saccade
lengths as well as the distribution of saccade durations are
skewed to the right, and are also well described by gamma
distributions. The temporal dependence between fixation durations
is very weak, as judged by the autocorrelation functions and ARIMA
time series fitting.

\subsection{Temporal analysis}
Art is considered to be a subjective field. How one views artwork
is individually unique, and reflects one's experience, knowledge,
preference, and emotions. According to Locher's two-stage model
\citep{locher}, exploration of a picture starts with a global
survey of the pictorial field in order to get an initial overall
impression of the structural arrangement and semantic meaning of
the composition. The second phase of an aesthetic episode consists
of visual scrutiny or focal analysis of interesting pictorial
features detected initially to satisfy cognitive curiosity and to
develop aesthetic appreciation of the display. Some pioneering
investigations into visual exploratory behavior of paintings by
\cite{buswell} and \cite*{yarbus}, and subsequent studies on the
informative details of an image by \cite*{antes} and
\cite*{mackworth} revealed that observers focus their gaze on
specific areas of the image, rather than in a random fashion. The
areas receiving high intensities of fixations were interpreted as
guiding the observer's interest in informative elements of the
image \citep*{henderson}.

\begin{figure}
\centering
 \includegraphics[width=1\textwidth]{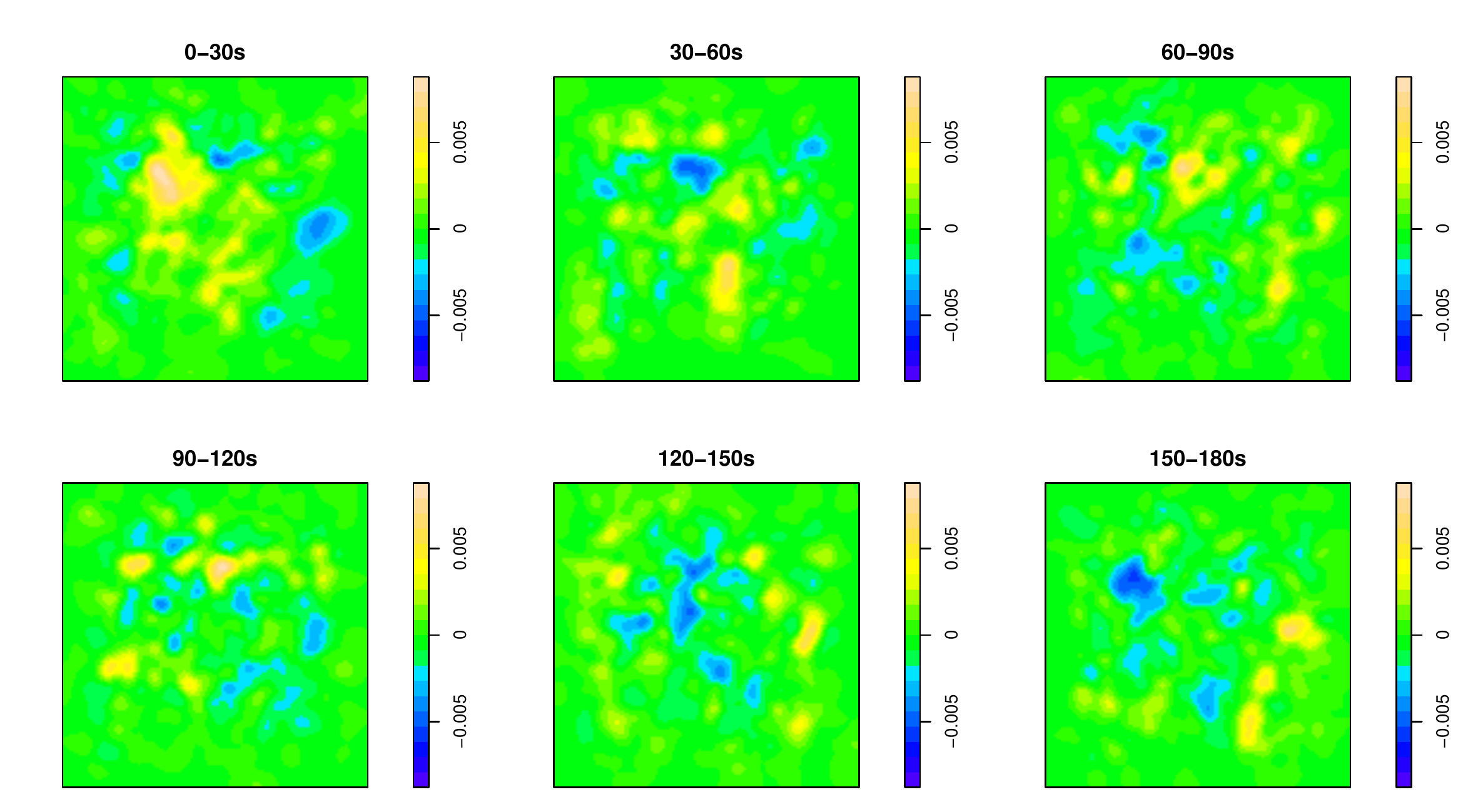}\\
  \caption{Residual intensity surfaces in 30s intervals. The more yellow the color, the higher the intensity compared to the mean intensity surface; the more blue the color, the lower the intensity.}
  \label{fig:Intensities_30s}
\end{figure}

Since people do not look at all parts of the painting, but rather
focus on some features of interest, the fixation process cannot be
assumed to be spatially stationary. Furthermore, the features
people look at may vary in time. To investigate the latter, we
estimated the intensity surface of fixation locations based on the
fixations of all subjects, and plotted residual intensities in 30s
intervals (Figure \ref{fig:Intensities_30s}). The residual
intensity is the difference between the estimated intensity
surface and the mean intensity surface of the six intervals. The
intensity in location $x\in I$, where $I$ is the area of the
painting, is estimated using the edge-corrected kernel estimator
(see e.g.\ \cite*{gbdr})
\begin{equation}
\label{eq:intens} \hat{\lambda}_h(x) = \dfrac{\sum_{i=1}^n
h^{-2}\, K(h^{-1}(x-{x_i}))}{\int_{I} h^{-2}
K(h^{-1}(x-u))\, du},
\end{equation}
where {$x_i$}, $i=1,...,n$, are the fixation locations, $h$ a
bandwidth (a smoothing parameter) and $K$ a kernel function, here
the standard two-dimensional Gaussian density. The bandwidth for
the mean intensity estimation is selected by applying the
cross-validation approach described in \cite*{diggle} and
\cite*{bermandiggle} and that bandwidth (here 17) is the same for
all intervals.

One can notice in Figure
\ref{fig:Intensities_30s} that during the first 30s, the gaze is
more concentrated on the tall pine tree on the left hand side of
the painting and after that more on the small tree in the middle
than on average during the whole three minutes time period. During
the second and the third minutes fixations are spread out a little
more than during the first minute. In addition, during the last
minute the areas which were of interest at the beginning (the two
trees) are now being avoided. Note that here we have only made
some visual observations. For a more formal comparison of
intensity surfaces we could use the test we introduce in Section
\ref{sec:comparison} below.

\begin{figure}
\centering
  \includegraphics[width=1\textwidth]{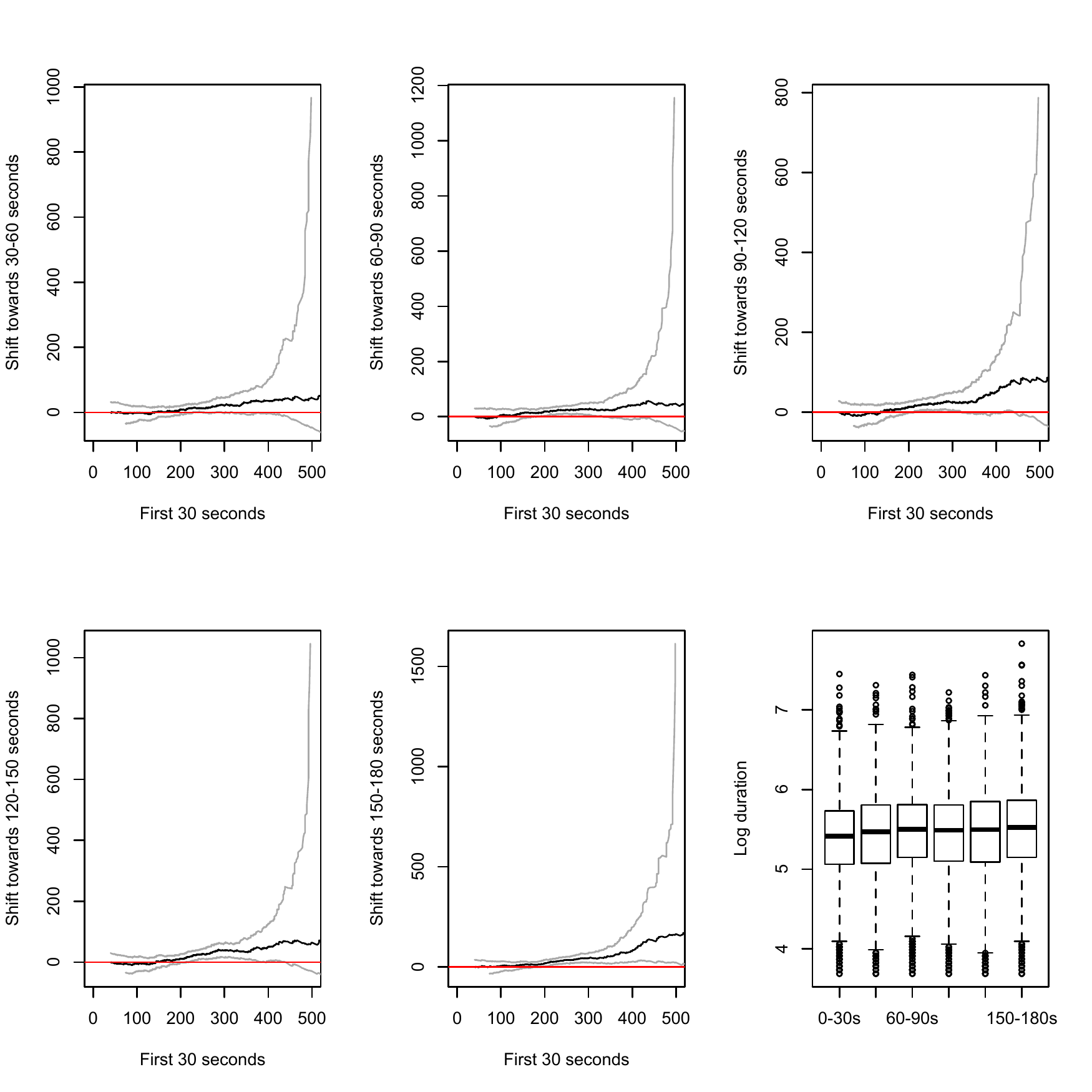}\\
  \caption{Fixation duration shift function estimates in 30s intervals, relative to the first 30s. The red line corresponds to equal distributions, and the grey lines form a 95\% asymptotic simultaneous confidence bands. The lower right panel is a box plot comparison of the distributions on a log scale.}
  \label{fig:shiftbytime}
\end{figure}

We also investigate the difference in fixation duration
distributions between the different 30s time intervals. Figure
\ref{fig:shiftbytime} shows a graphical way to compare
distributions, a shift function \citep{doksumsievers} {defined as follows}. If $X$ has
the cumulative distribution function $F$, written $ X \sim F$, and {$Y$ the cumulative distribution function $G$,} $Y\sim G$, the shift function is defined as $$\Delta (x) =
G^{-1}(F(x))-x, $$ and has the property that $X+\Delta (X) \sim
G$. The idea of the shift function is to compare two distributions
based on how much one distribution function needs to be shifted in
order to coincide with the other. In general, this may depend on
the abscissa at which the comparison is made. {We have the simplest case when
the second distribution is simply} the first shifted by a
constant amount. The shift function is then just a horizontal line
at the level of the shift (in particular, if the distributions are
the same, the shift function is a horizontal line at the level
zero). If a location-scale model is appropriate, the shift
function is a straight line with slope related to the scale
change. The shift function is easily estimated using empirical
distribution functions, and simultaneous confidence intervals
based on the distribution of the Kolmogorov-Smirnov statistic are
given in \cite{doksumsievers}. If a horizontal line at level 0
falls inside the band, the two distributions are statistically
indistinguishable. The shift function estimates in Figure
\ref{fig:shiftbytime} represent the comparison of the second
through the sixth 30s fixation distributions to the first 30s
distribution. It seems that the later time intervals tend to have
slightly longer fixations between about 200ms and 450ms than the earlier intervals. This is
also illustrated by the side-by-side box plots in the lower right
hand panel of the figure.

\section{Comparison of novices and non-novices}
\label{sec:comparison}

In the previous section we introduced some tools to describe the
fixation process. Now these tools are applied to compare the
fixation processes of novices and non-novices. First, we compare
the fixation intensity surfaces of the two groups visually and
then, construct a test for more formal comparison. Second, we
compare the fixation duration distributions in the two groups
using shift plots. Finally, we investigate whether the fixation
duration distributions change within the two groups if the
painting is replaced by another.

\subsection{Visual comparison of intensity surfaces}

\begin{figure}
\centering
  \includegraphics[width=0.7\textwidth]{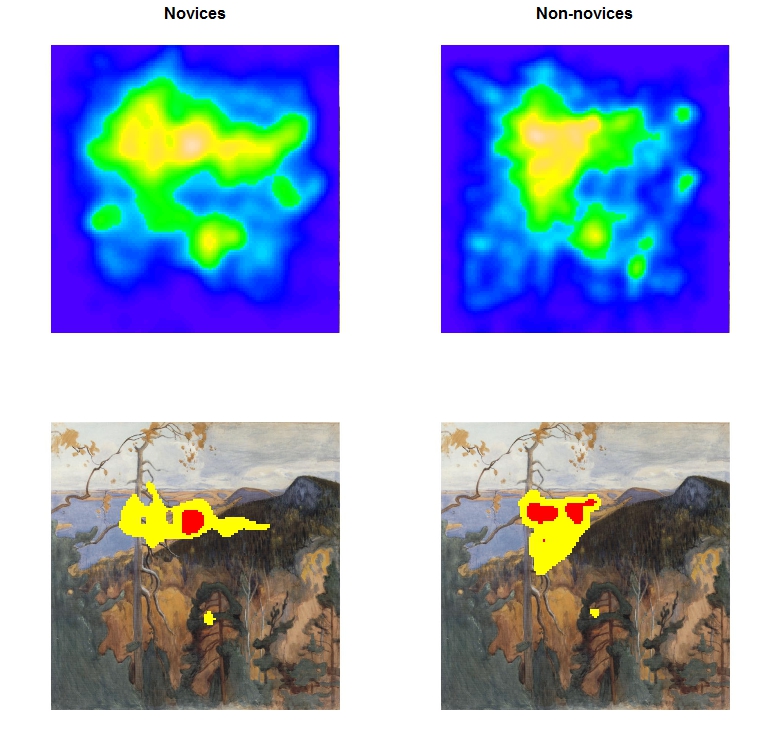}
  \caption{Top row: The estimated overall intensity surface with bandwidth 20 for novices (on the left) and with bandwidth 16 for non-novices (on the right). Bottom row: Top 5 \% (yellow) and top 1 \% (red) intensities, novices vs. non-novices.}
  \label{fig:Top5_nov_non}
\end{figure}

To compare the spatial patterns of fixations, we first estimate
the overall intensity for each group by using equation
(\ref{eq:intens}), see the top row in Figure
\ref{fig:Top5_nov_non}. We see that both groups concentrate on the
middle part of the painting and do not look closely at the edges,
and that the fixations of non-novices seem to be slightly more
concentrated than the fixations of novices. Non-novices look a lot
at the large pine and one of its branches (vertical alignment)
while novices look at the same branch of the tree but continue to
the right of the branch (horizontal alignment). We also plot the
areas of the painting that are visited most (top 5\% in yellow and
top 1\% in red) for each group (Figure \ref{fig:Top5_nov_non},
bottom row). This verifies the finding of non-novices' interest in
the vertical trunk while novices have been concentrating more on
the horizontal branch.

A more detailed investigation, where the time interval is divided
into 30s intervals (see supplemental article \cite*{suppC}), reveals that especially early on
the non-novices are more concentrated on the large tree and its
branch than the novices. The gaze of novices starts to spread out
much earlier than the gaze of non-novices and novices have more
hot spots than non-novices. The gaze of non-novices mainly stays
on the large pine tree and in the end also on the small tree on
the right side of the large one, while novices investigate other
areas of the painting as well.

\subsection{Test for comparing two intensity surfaces}
\label{sec:comparisonintensities}

In order to test whether two intensity surfaces differ
significantly from each other, i.e.\ whether the subjects in the
two groups have been looking at different areas of the painting,
we apply the method developed by \cite*{kelsalldiggle95a,
kelsalldiggle95b}. They considered a ratio of two kernel density
estimates and used a Monte Carlo test for testing if the ratio is
constant, i.e.\ whether the kernel estimated intensity surfaces
can be considered similar. This ratio is used in epidemiology to
explore whether some diseased cases are randomly located among
healthy controls and is therefore, often called a relative risk
function. The logarithmic relative risk function is defined as
\(\rho(x) = \text{log}({\lambda_1(x)} / {\lambda_2(x)})\), where 
$\lambda_i$'s are computed using equation (\ref{eq:intens}), and
can reveal areas where the intensity of fixations differs between
novices and non-novices.

By conditioning on the observed number of fixations, we may view
the fixation locations of novices and non-novices as independent
samples from the densities $f_1(x)$ and $f_2(x)$, where
\[\hat{f}_i(x) = \frac{\hat{\lambda}_i(x)}{\int_{I} \hat{\lambda}_i(x) dx},\]
for {$i=1,2$}. We can now define a logarithmic density ratio
\[r(x) = \text{log}\ \frac{\hat{f_1}(x)}{\hat{f_2}(x)} = \rho(x) \ - c,\]
where \(c = \text{log}(\frac{\int_{I} \hat{\lambda}_1(x) dx}{\int_{I}
\hat{\lambda}_2(x) dx})\) (see e.g. \cite*{wakefield},
\cite{kelsalldiggle95b}). {We have}
\[\hat{r}(x) = \text{log}\ \frac{\hat{f_1}(x)}{\hat{f_2}(x)},\]
which contains all the information about the spatial variation in
$\rho(x)$.

According to \cite{kelsalldiggle95b}, the choice of the kernel
function is not critical, but choosing the smoothing parameter $h$
is. Furthermore, it is not obvious whether one should use the same
bandwidth for both densities when computing the ratio.
\cite{kelsalldiggle95b} suggest that if the densities are expected
to be nearly equal, the bandwidths should be equal as well to
reduce bias. They also warn that if the sample sizes are very
different, one can get poor results when using the same bandwidth
for both samples. Furthermore, \cite*{baileygatrell} suggest that
the kernel estimate in the denominator should be deliberately
oversmoothed (a larger bandwidth should be used). In addition,
\cite*{hazelton} shows that edge correction terms are not needed
if a common bandwidth is used.

We test the null hypothesis that the intensity surfaces are equal,
i.e.\ that the logarithmic density ratio \(r(x) = 0\), by the
following Monte Carlo test \citep*{barnard}. Ten subjects were
randomly selected from the twenty subjects to form the novice
group and the remaining ten subjects formed the non-novice group.
(Hence the possible dependence between the fixation locations
within a subject was taken into account.) By doing this random
grouping $m$ times, we obtain $m$ log-density surfaces
\(\hat{r}_1, \dots, \hat{r}_m\), under the null hypothesis. As
introduced by \cite{kelsalldiggle95b}, we use the test statistic
\begin{align*}
T_j = \int_I { \hat{r}_j(x)^2} dx, \quad j = 0,
\dots, m.
\end{align*}
The $p$-value is then \(p = \frac{k+1}{m+1}\), where $k$ is the
number of test statistics $T_j$ which are larger than or equal to
the observed value of the test statistic, $T_0$.

Since the number of fixations vary quite a lot between the
subjects, we estimate the bandwidths separately for the two groups
by using the cross-validation method by \cite{diggle} and
\cite{bermandiggle} with Ripley's isotropic edge correction
\citep*{ripley}. The estimated bandwidth for the novices was
approximately 20 and for the non-novices 16. We obtained $T_0 =
59604$ and $p = 0.108$ after random grouping of subjects 10000
times. This means that the difference between the intensity surfaces of novices and non-novices is not statistically
significant even though there are some dissimilarities in the
intensities, see Figure \ref{fig:Top5_nov_non}. The largest differences between the intensities of
the two groups are near the edges of the painting, where there are
very few fixations of either group. 

We made the same comparisons for the other five paintings and obtained similar results, see supplemental article \cite*{suppD}. None of the tests gave any significant result, and Fisher's combined probability test resulted in $p = 0.392$ ($\chi^2 = 12.685$, $df = 12$). Therefore, we did not find any overall differences when comparing the intensities of the two groups. However, it seems that the more abstract paintings could reveal some differences between novices and non-novices, see results related to Kandinsky's painting in \cite{suppD}.

\subsection{Comparing duration distributions}

\begin{figure}
\centering
  \includegraphics[width=0.5\textwidth]{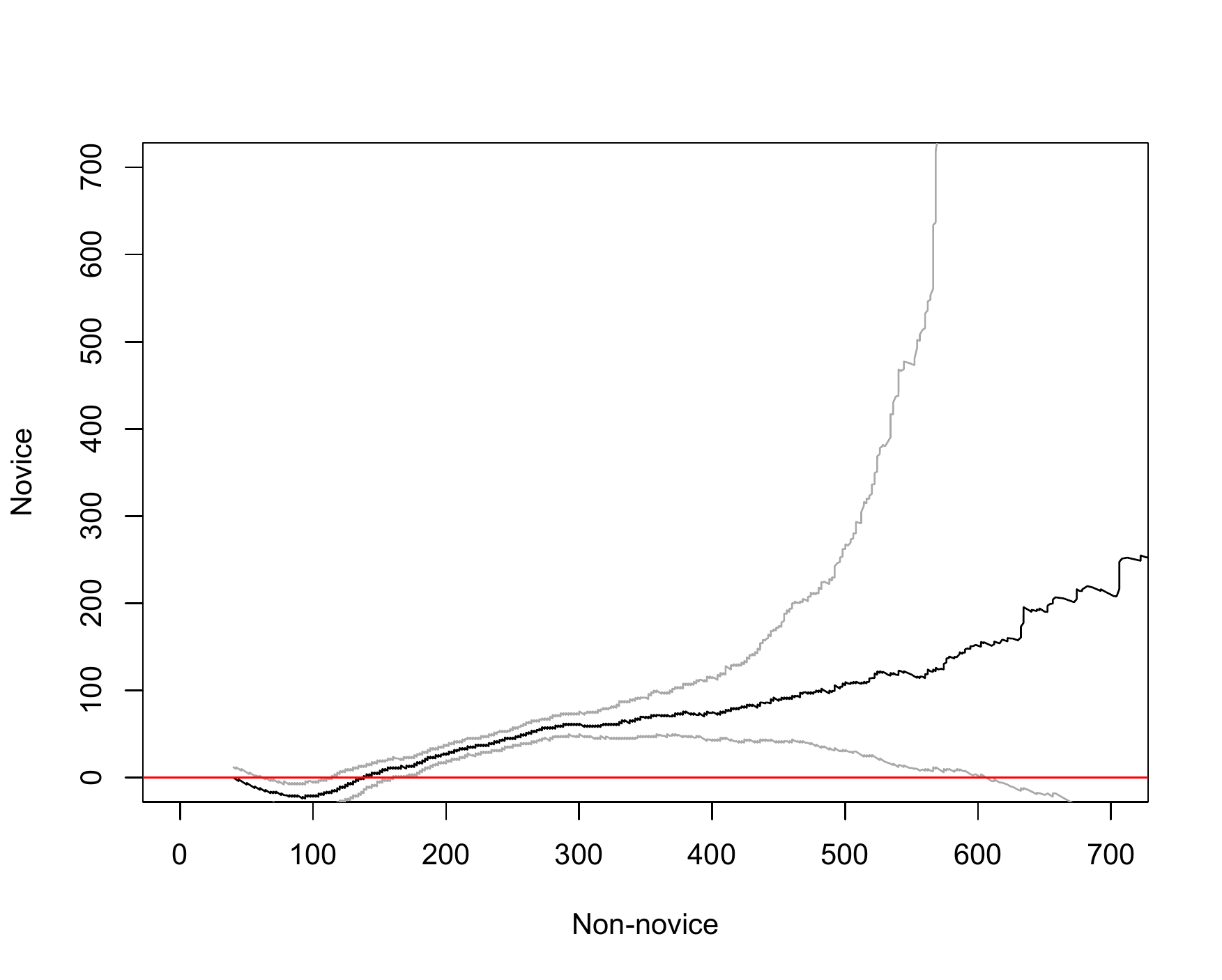}\\
  \caption{Shift plot of non-novice distribution vs. novice distribution for fixation duration. The red line indicates equal distributions, and the grey lines form a simultaneous asymptotic 95\% confidence band.}
  \label{fig:shiftfixartnonart}
\end{figure}

To compare the fixation duration distributions of the novice and non-novice groups,
we again look at the shift plot, for the J\"arnefelt painting see Figure \ref{fig:shiftfixartnonart}.
Here, the novices have fewer short fixations (duration less
than 125 ms), and more fixations between about 175 and 600 ms than
non-novices. Similar results were found for the remaining five paintings, see \cite{suppD}. This is in line with the early findings of
\cite{buswell}, and may mean that novices more rarely only glance
a point on the painting but rather spend more time on each point
compared to non-novices. 

\subsection{Pairwise comparison of paintings}

\begin{figure}
\centering
  \includegraphics[width=1\textwidth]{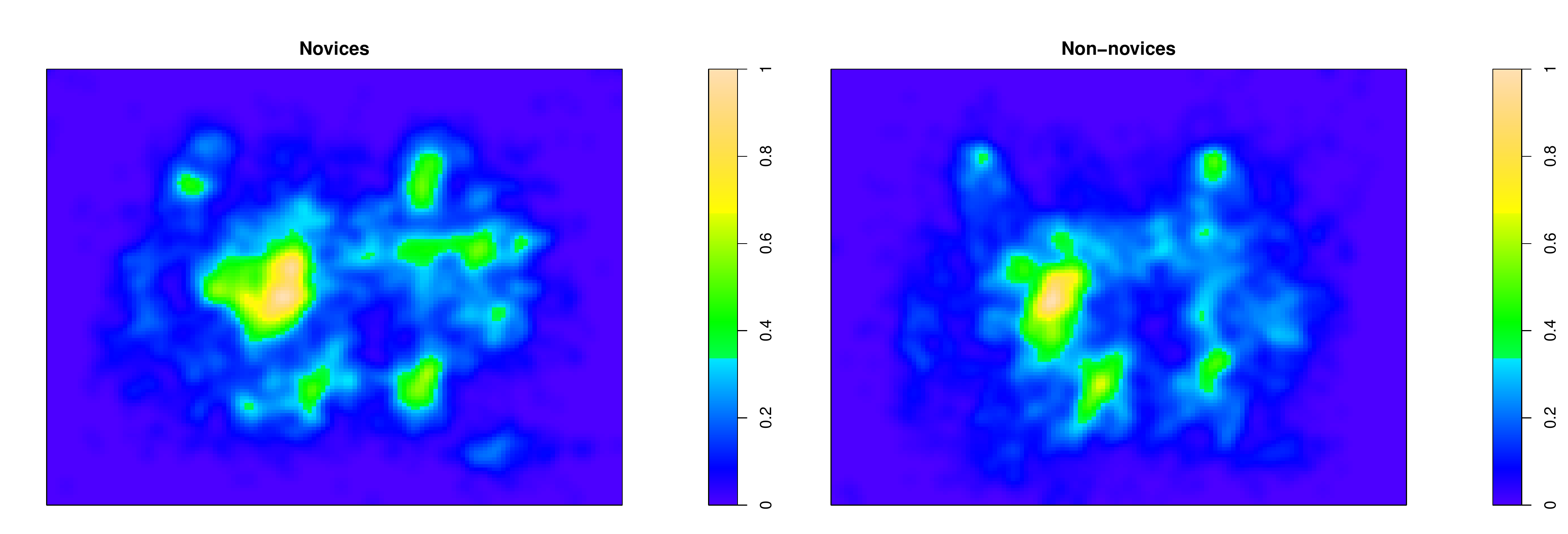}\\
  \caption{The estimated overall intensity surface with fixed bandwidth 13 for novices (on the left) and for non-novices (on the right) for the Monet painting.}
  \label{fig:Monet_intensity}
\end{figure}

Since several paintings are available, we are also able
to investigate whether the fixation
process is similar for different paintings, i.e.\ whether the
fixation process stays stable when looking at paintings. We are
particularly interested in whether we detect any differences
between the behavior in the novice and non-novice groups. We
started by comparing the J\"arnefelt painting and the painting Terrace at Sainte-Adresse by Claude Monet
(1867) shown in Figure \ref{fig:paintings}(b) since both are landscapes and hence similar to each other. The estimated intensity surface of the fixation
locations for Monet's painting is also shown separately for novices and non-novices in
Figure \ref{fig:Monet_intensity}. The intensity surfaces of the
two groups look fairly similar, see also the formal test in \cite{suppD}.

\begin{figure}
\centering
  \includegraphics[width=0.5\textwidth]{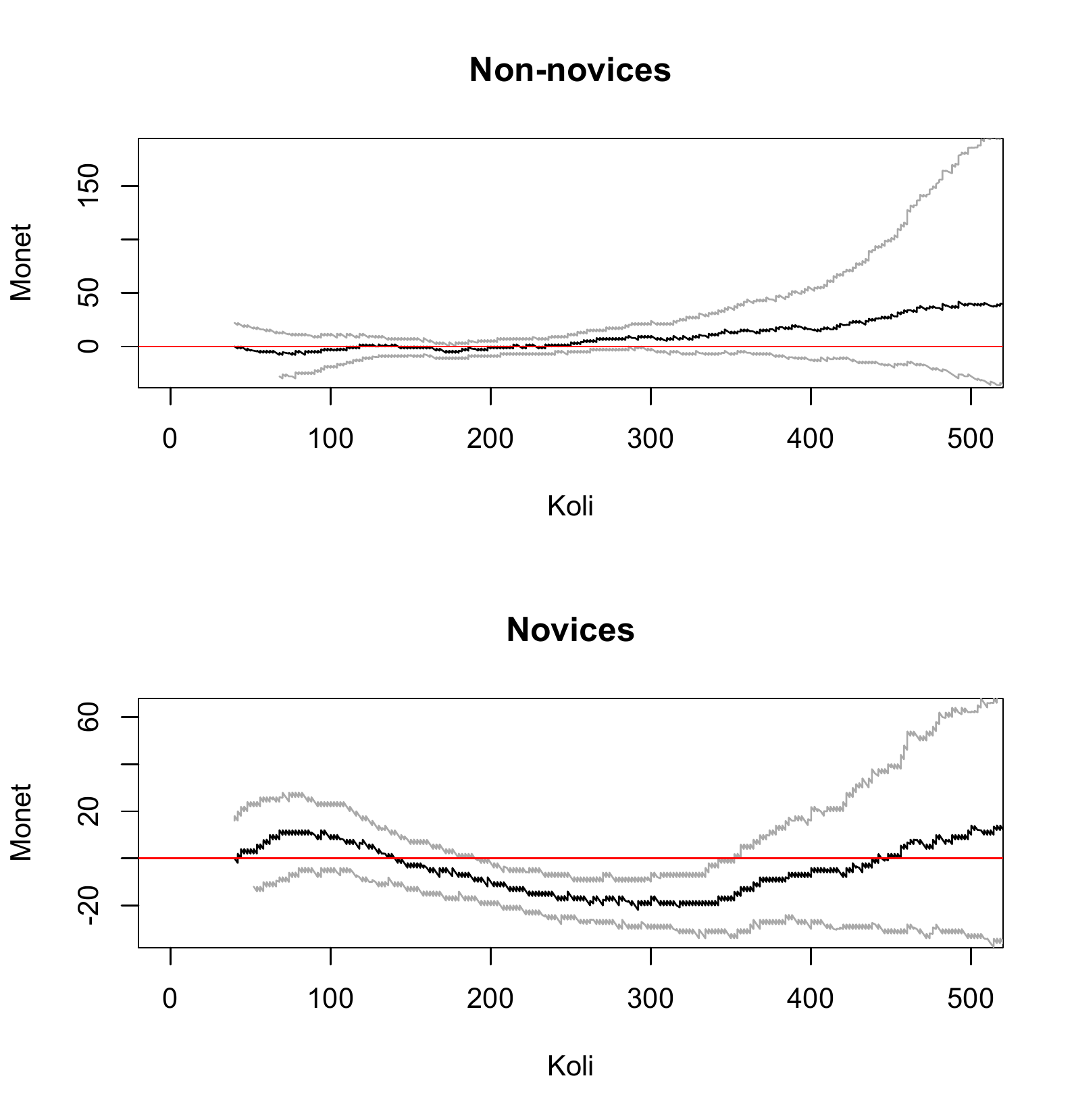}\\
  \caption{Shift function comparison of fixation duration distributions between the J\"arnefelt and Monet paintings for non-novices (upper panel) and novices (lower panel). The red lines correspond to equal distributions and the grey lines form 95\% asymptotic simultaneous confidence bands.}
  \label{fig:monetkolidurations}
\end{figure}

Since the pictures are completely different, the intensity
surfaces of fixations for the paintings are also different and comparing them would be meaningless. However, it is of interest to compare the
fixation duration distributions of novices and non-novices between two paintings in order to see whether the way of looking at
paintings remains stable between different paintings. The shift
function comparisons of the duration distributions for novices and
non-novices are plotted in Figure \ref{fig:monetkolidurations} for the J\"arnefelt - Monet pair. We
can see that for non-novices there is no significant difference
between the paintings, while for novices there are more fixations
that last 200-350 ms on the J\"arnefelt painting than on the Monet
painting. We made similar comparisons for the other pairs of paintings as well, see supplemental article \cite*{suppE}. Here, only the results based on the comparisons of the fixation duration distributions for the most similar pairs of paintings, namely Suomi and Kandinsky which are abstract paintings (see Figure \ref{fig:paintings}(c) and (d)), and Poussin and Tammi which have people in focus (see Figure \ref{fig:paintings}(e) and (f)), are included. For novices, we see a difference for each three pairs of paintings. For non-novices, however, we can only see a clear difference for the Suomi - Kandinsky pair, but there is no difference for the other two pairs of paintings.
This indicates that the non-novices, being used to
looking at paintings, are more consistent in their fixation durations
between paintings, but for novices, the duration of fixations
depends on the painting.

\section{Stochastic model}
\label{sec:model}

Above, we used intensity surfaces and shift plots to compare the
eye movements of novices and non-novices. Next, we will construct
a simple spatio-temporal model that describes the dynamics of the fixation process. We
will fit the model separately to the novice and non-novice groups for the J\"arnefelt painting
and see if we can find some further differences between the groups
by comparing the fitted models in terms of functional summary statistics and their model based simultaneous envelopes.

To build a spatio-temporal model for the fixation process, we need
to model fixation locations, fixation durations, saccade durations
and saccade lengths. (Saccade durations are needed to fill the
whole three minutes time period.) As building blocks we use the
tools discussed in Section \ref{sec:data}. Fixation locations are
modeled as a realization of a spatial point process, which is
characterized by its intensity function. Furthermore, appropriate
distributions can be fitted to the fixation and saccade durations
and for saccade lengths. After having estimated the intensity
surface and the distributions mentioned above, we can simulate
from the resulting model and compare the behavior of the model to
the behavior in the data, and finally compare the models fitted to
the two groups.

A natural reference or null model for spatial point pattern data,
locations of fixations in our case, is the homogeneous Poisson
process, where the points are located uniformly and independently
of each other on the study area. In our situation, however, only
the more interesting parts of the painting are visited, and we
cannot assume that the final set of locations of fixations is a
realization of a homogeneous process. A homogeneous Poisson
process is therefore not a realistic reference process for the
fixation locations. Next, we define our own
reference model for the fixation process.

\begin{figure}
\centering
  \includegraphics[width=0.5\textwidth]{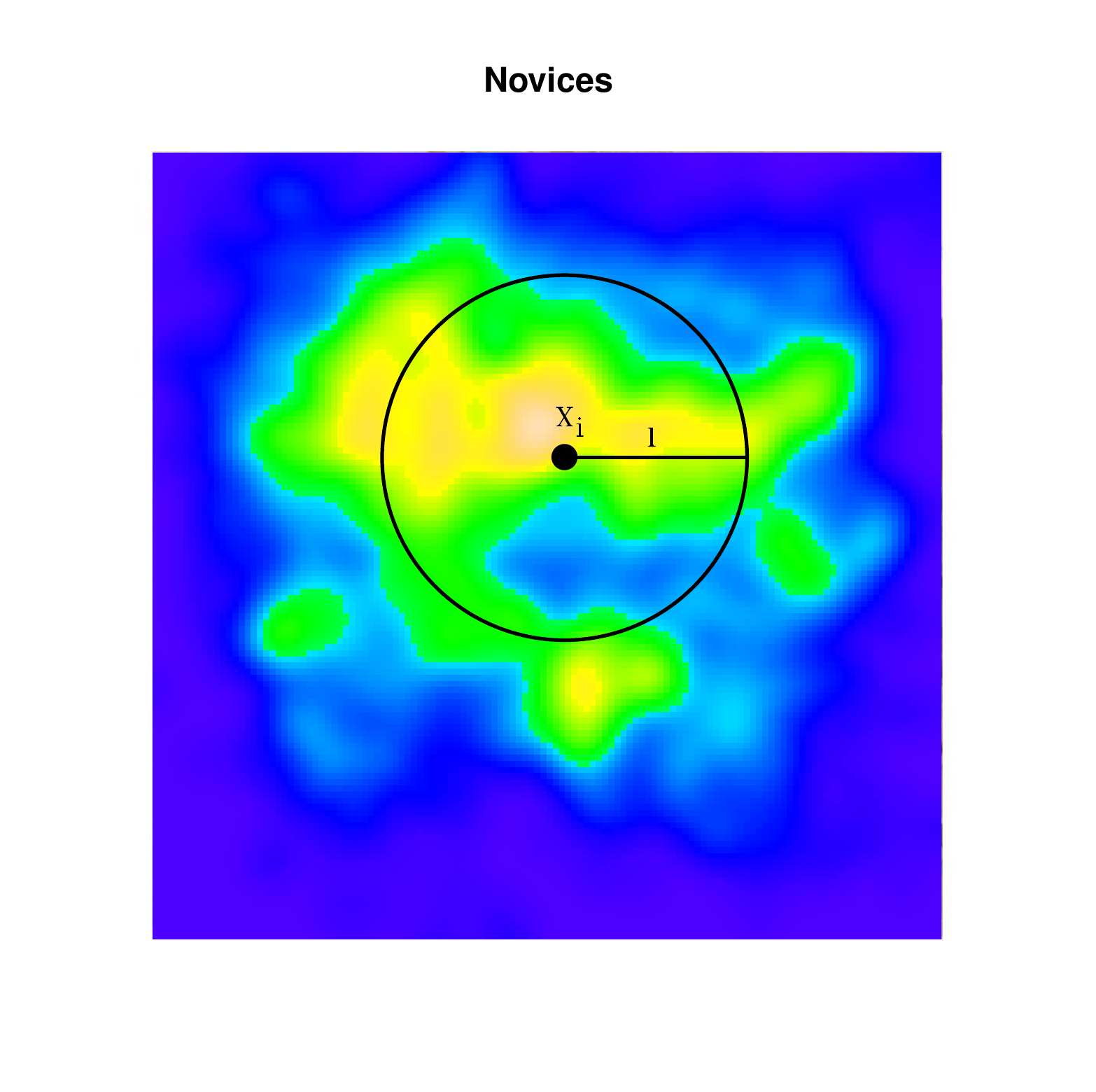}\\
  \caption{Next fixation location is chosen from the black circle at distance $l$ from the current fixation $x_i$ according to the estimated intensity surface.}
  \label{fig:Next_fixation}
\end{figure}

Given an intensity surface, the first fixation
location is drawn from the probability distribution on the
observation window by normalizing the intensity to a bivariate
distribution. The length $l$ of the next saccade is drawn from a
saccade length distribution for the group in question. Then, the
exact location of the new fixation is decided by taking a point at
distance $l$ away from the current fixation location according to
the intensity surface. In other words, the new location
is chosen from the conditional intensity of points at the given
distance from the current location, see Figure
\ref{fig:Next_fixation}.

Concerning the temporal evolution, it seems natural to draw each
fixation duration independently from the distribution of
durations.

\section{Model fit and comparison between novices and non-novices for the J\"arnefelt painting}
\label{sec:results}

In what follows, we will first give the particular choices of the distributions mentioned in the previous section, and introduce some summary statistics that could be used to assess the goodness-of-fit of the fitted models. Finally, we fit the model to the data for the J\"arnefelt painting,
separately for novices and non-novices, simulate 200
realizations of each process, and check the goodness-of-fit of the fitted models by comparing the summary statistics estimated from the
data to those estimated from the simulations.

\subsection{Choice of distributions}
The intensity surfaces of the fixation locations are estimated
separately for each group and the bandwidth is selected by
applying the cross-validation approach implemented in the R package \verb+spatstat+ \citep{baddeley}. The same bandwidth is used
for the data and for the simulations.
The first location
is drawn from the intensity surface estimated from the locations
of the first fixations of the subjects in the group.
Alternatively, the intensity surface estimated based on all the fixation locations in the group could be used. We believe that the choice of the first fixation location is not crucial.

Given the summary statistics computed from the data (Section \ref{sec:datadescription}),
an appropriate distribution for the duration of  fixations and
duration of saccades is the gamma distribution. Therefore, we
model the fixation durations by sampling from the estimated gamma
distribution of fixation durations in the group. The duration of each saccade is
modeled by drawing from the common distribution of saccade
durations estimated from all the subjects, as saccades are
involuntary, once started, and therefore not expected to vary
depending on viewing experience. The intensity surface, as well as
the fixation duration and saccade duration
distributions, stay the same throughout the observation period.

For saccade lengths we first fit a Gamma distribution separately for novices and non-novices and then, use truncated versions of these distributions in order to avoid jumps outside the painting. The truncation point depends on the current location: if the process is at location $x_i$, the longest jump it can take is to the furthest corner of the painting. This distance is called $l_{i,max}$ and serves as the truncation point for the truncated gamma distribution from which the next jump is sampled. We noticed, however, that the truncated gamma distribution alone does not catch the long jumps that the gaze makes when the person moves her/his attention to another location. Therefore, we include such long jumps into the model and define the distribution for saccade lengths to be a mixture of two distributions: When the process is at location $x_i$, the length of the next jump is sampled from the uniform distribution $U(l_{i,max}/2,l_{i,max})$ with probability $p$ and from the truncated Gamma distribution with truncation point $l_{i,max}$ with probability $1-p$. After some experimentation, we fixed the probability $p$ for sampling the saccade length from the uniform distribution above to 0.2 for both groups. 

\subsection{Summary statistics}

To get some idea of how the gaze jumps between different areas of
the painting in the data and in the simulations, we divide the
painting into quarters and compute the transition frequencies
between them. Each quarter is called a state and identified with
numbers 1--4 (from upper left to lower right quarter).

In order to see how much of the painting is viewed, we use two
functional summary statistics, namely convex hull coverage and
ball union coverage defined as follows. Let, as before, $I$
represent the observation window and $x_i \in I$ be the $i$th
fixation of the fixation process $\{X(t)\}$. An ordered set of
$n_t$ fixations up to a fixed time $t$ is denoted by
$x=\{x_1,\dots,x_{n_t}\}$. The convex hull of this point set is
\[
C_x(t)=\left\{\sum_{i=1}^{n_t}\alpha_ix_i\,:\, \alpha_i\geq 0
\mbox{ for all $i$ and $\sum_{i=1}^{n_t}\alpha_i=1$}\right\}\,
\]
and the relative area of the convex hull is denoted by
\[
AC_x(t)= \left\{
\begin{array}{ll}
0&\mbox{ if $n_t<3$}\\
\frac{{\rm area}(C_x(t))}{{\rm area}(I)}&\mbox{ if $n_t\geq
3$}.\,
\end{array}
\right.
\]
This is called the convex hull coverage of the fixation process.
Hence the convex hull coverage is computed every time a new
fixation appears.

Let $b(x_i,R)$ be a ball of fixed radius $R$ centered in a
fixation $x_i \in I$. Define
\[
U_x(t)=\{\cup_{i=1}^{n_t}b(x_i,R)\cap W\},
\]
{and call
\[
AU_x(t)=\frac{{\rm area}(U_x(t))}{{\rm area}(I)}
\]
the} ball union coverage of the fixation process.

In addition, to measure the total length of the distance the gaze
has moved, we use the scanpath length function \citep*{noton},
which is the sum of the saccade lengths up to time $t$. Hence, the
scanpath length can be defined as
\[
L(t) = \sum_{i=1}^{n_t-1} l_i  \mathbf{1}(t_{i+1} \le t),
\]
where $t_{i+1}$ is the time when the fixation $i+1$ at location
$x_{i+1}$ takes place and $l_{i}$ is the length of the jump just
before that fixation. That length is the Euclidean distance between
two successive fixations, i.e.\ the length of the saccade.

\subsection{Model fit}

\begin{figure}
\centering
  \includegraphics[width=1\textwidth]{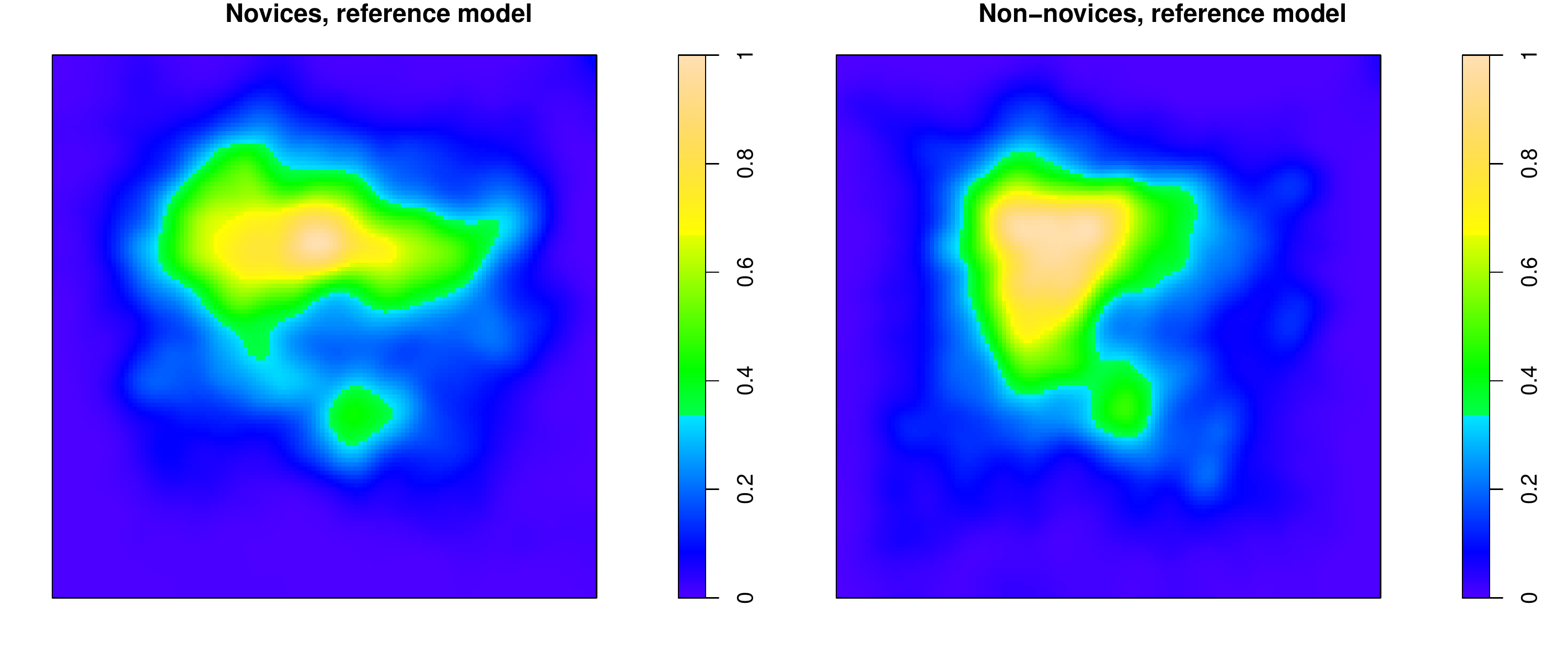}\\
  \caption{Scaled overall intensity surfaces of the reference model for novices (left) and for non-novices (right).}
  \label{fig:Intensities_model}
\end{figure}

The estimated model based intensity surfaces for the groups are different, see
Figure \ref{fig:Intensities_model}. As mentioned earlier, both
groups look a lot at the large pine tree and the top of the small
tree next to it. The novice group tends to follow the branch of
the large tree while the non-novice group follows the trunk.

\begin{figure}
\centering
  \includegraphics[width=1\textwidth]{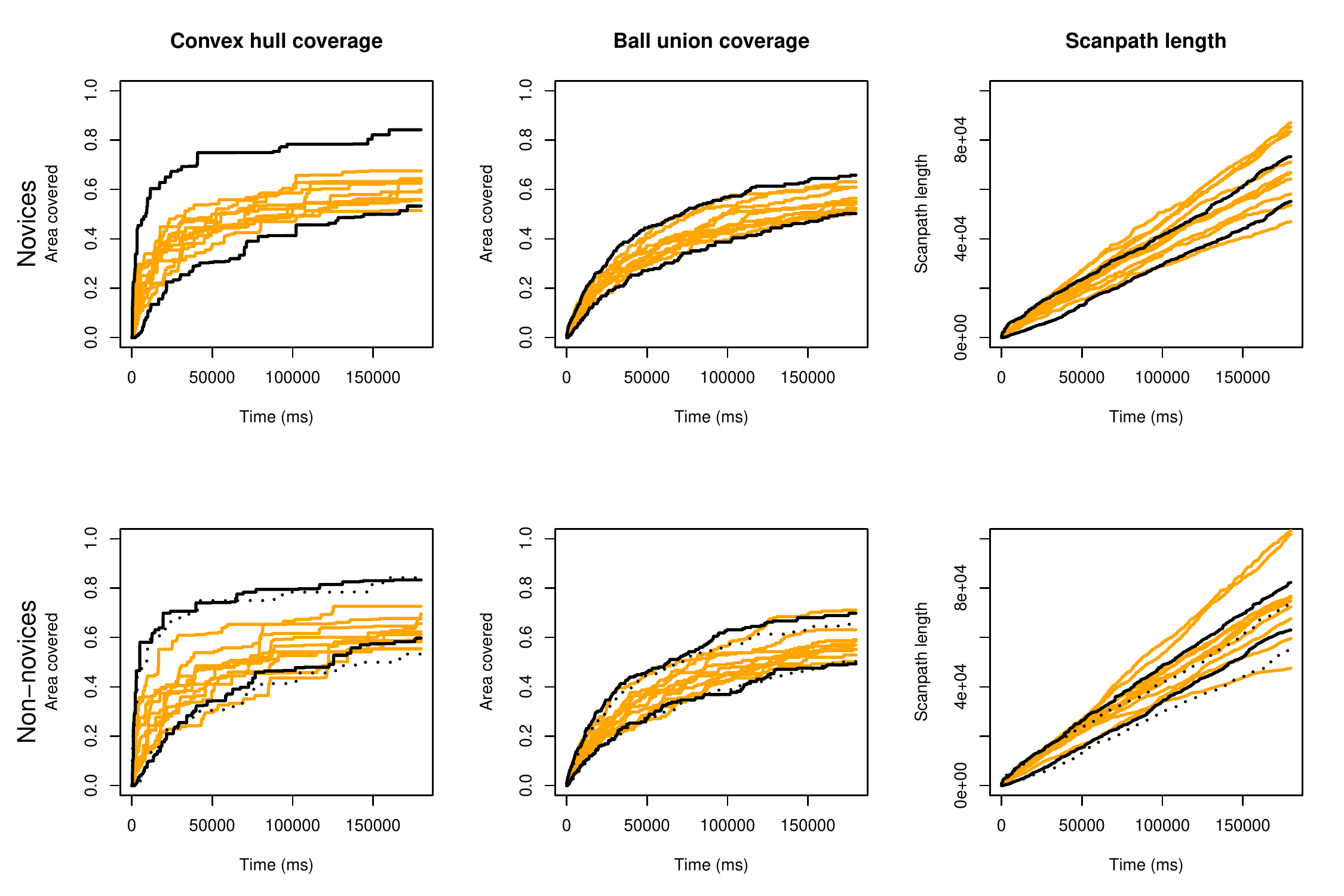}\\
  \caption{Convex hull coverage (left), ball union coverage (middle) and scanpath length (right) curves for novices (top) and non-novices (bottom). Orange lines represent the subjects and black solid lines represent 95\% simultaneous envelopes estimated from 200 simulated realizations of the reference model of the group in question. Dotted lines (bottom) represent 95\% simultaneous envelopes for novices.}
  \label{fig:ModifiedModel_summaries_both}
\end{figure}

The convex hull coverage, ball union coverage with disc radius 35
pixels, and scanpath length for non-novices and novices are shown and compared
to the models in Figure \ref{fig:ModifiedModel_summaries_both}. The 95 \% simultaneous (rank) envelopes were created by using the R package \verb+spptest+ \citep{myllymaki2, myllymaki}. Figure \ref{fig:ModifiedModel_summaries_both}
shows that the coverage statistics based on the model describe the
data quite well. The model based scanpath length is, however, less variable and
somewhat shorter in the simulations than in the data. It seems that the model is not able to catch all variation related to the scanpath length. In the beginning (during the first 30 seconds for novices and during the first minute for non-novices), the model describes the data well but later on, variation in the data becomes too large to be caught by the model. One should also note that the scanpath length is strongly affected by the number of fixations and saccade lengths, which both vary quite a lot in our data, see supplemental articles \cite{suppA} and \cite{suppB}.

\begin{figure}
\centering
  \includegraphics[width=0.9\textwidth]{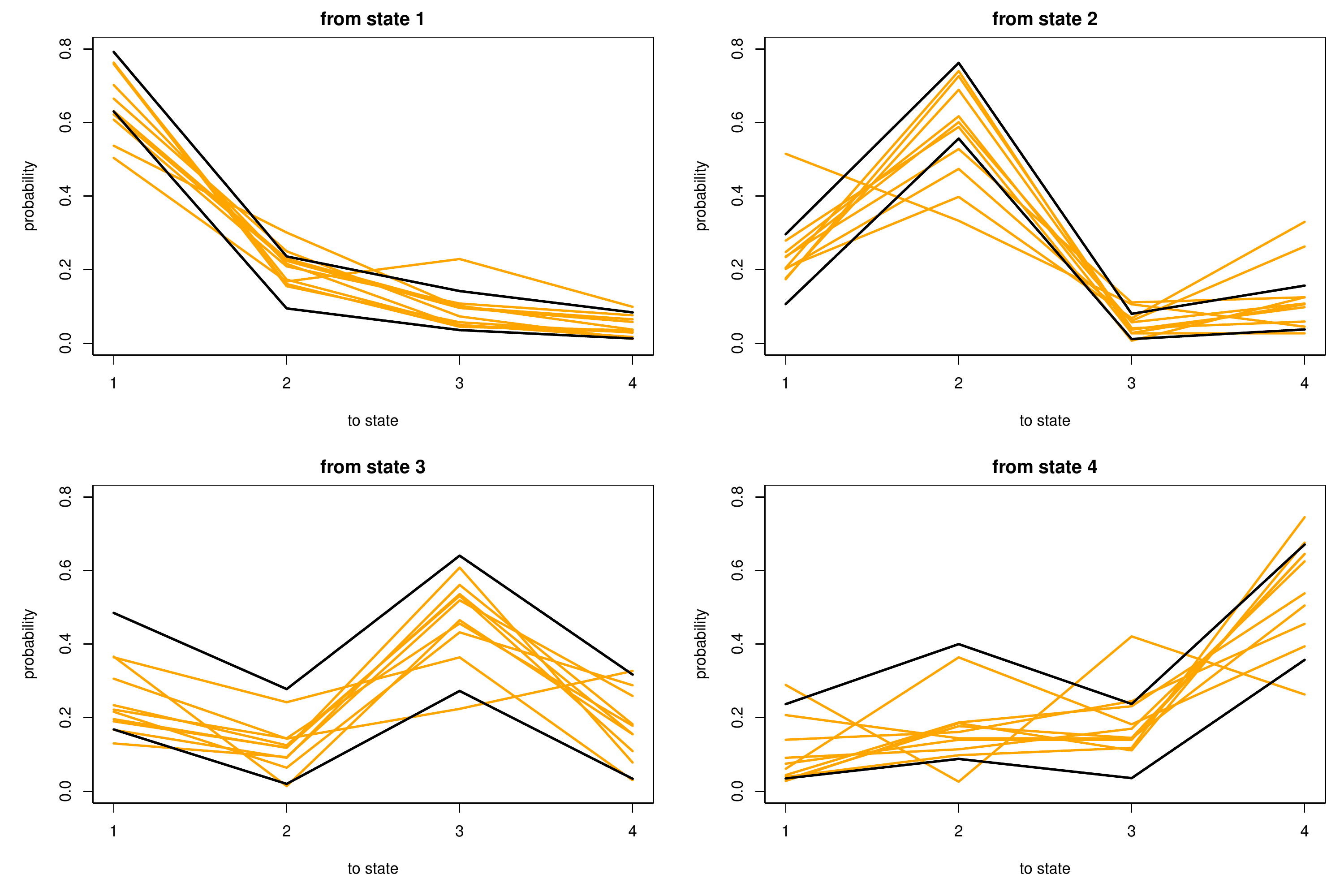}\\
  \caption{Transition probabilities for novices with 95\% simultaneous envelopes estimated from 200
    simulated realizations of the reference model.}
  \label{fig:ModifiedModel_tp_nov}
\end{figure}

\begin{figure}
\centering
  \includegraphics[width=0.9\textwidth]{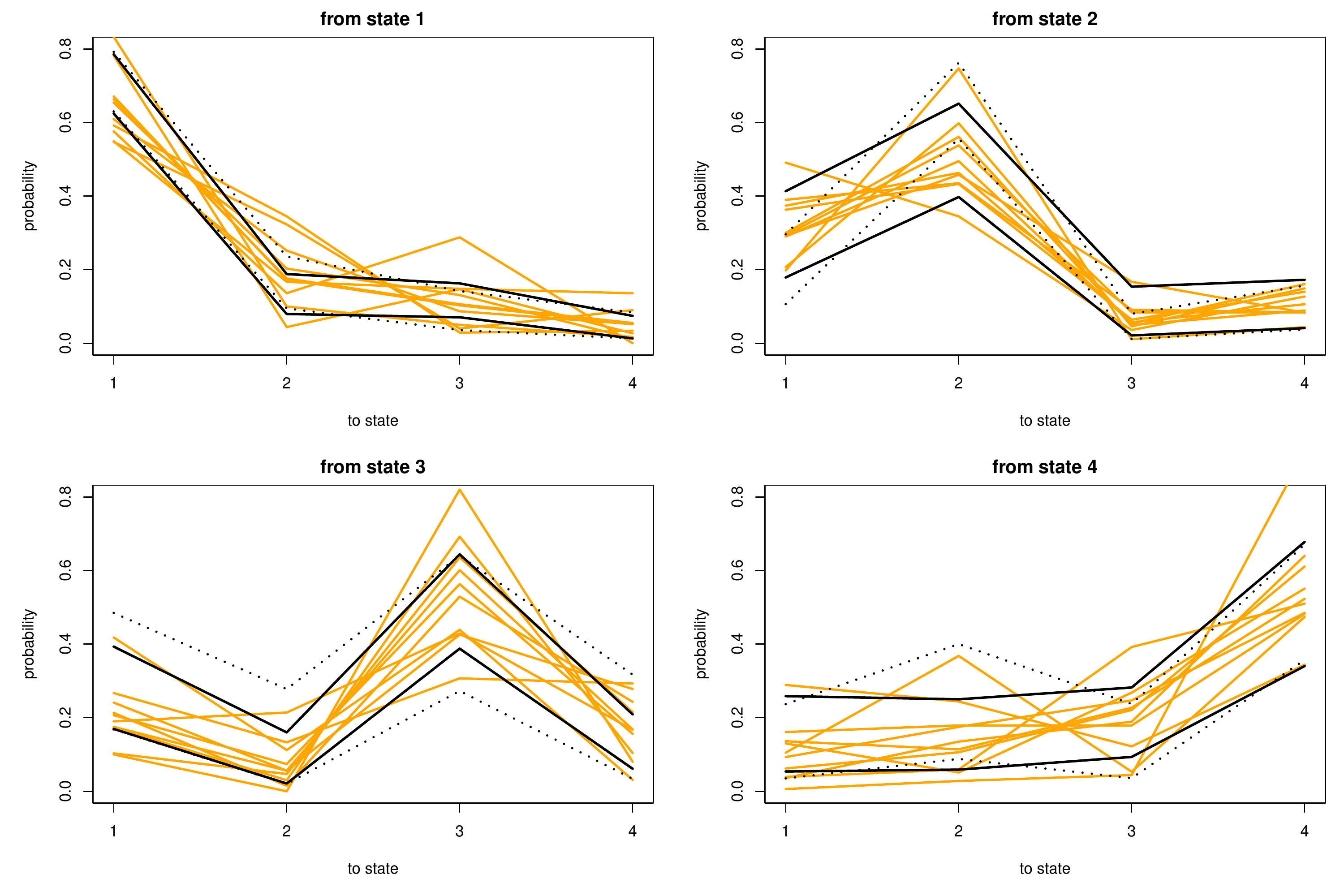}\\
  \caption{Transition probabilities for non-novices with 95\% simultaneous envelopes estimated from 200
    simulated realizations of the reference model. Dotted lines represent 95\% simultaneous envelopes for novices.}
  \label{fig:ModifiedModel_tp_non}
\end{figure}

For both the novice data and the non-novice data, the estimated transition probabilities between different quarters
of the painting indicate that the model does an adequate job
(Figure \ref{fig:ModifiedModel_tp_nov} and Figure \ref{fig:ModifiedModel_tp_non}). Recall that the quarters
of the painting are called states and numbered from 1 to 4 (from
upper left to lower right quarter). 

The model seems to fit quite well to the fixation process for both groups. The fitted models are quite similar but also some differences can be found. As can be seen in Figure \ref{fig:ModifiedModel_summaries_both},
the model based coverage summaries are quite similar for the two
groups. However, the modeled gaze of non-novices seems to move
slightly more than the gaze of novices according to the scanpath
length (see Figure \ref{fig:ModifiedModel_summaries_both}, bottom right plot). This is somewhat surprising, since the intensity surface of non-novices seems to be more compact than the one of novices,
which would indicate that the gaze of non-novices takes shorter
jumps in smaller areas than the gaze of novices. However, fixation
durations of novices are longer than the ones of non-novices in
general, which means that non-novices tend to make more jumps during
the three minute period compared to novices. Thus, it is possible
that the gaze of non-novices makes a larger number of short jumps than the gaze of novices which results in the gaze traveling a longer path than the gaze of
novices.

\section{Conclusions and Discussion}
\label{sec:disc}

In this study we have analyzed eye movement data from twenty
subjects who had looked at six different paintings. We have been
particularly interested in whether non-novices and novices look at paintings in the same way, both when it comes
to which parts of the painting they look at and how long the gaze
typically stays in one spot. By regarding the fixation process as
a spatio-temporal point process and by using a new set of tools we
can study the differences from a slightly different perspective
than what has been done earlier. Our main purpose has been to see
how useful our new approach is and therefore, we have mainly
concentrated on one of the six paintings that we have had
available, namely the J\"arnefelt painting {\em Koli landscape}. However, some results based on the other paintings are included.

By looking at the intensity surfaces of the point patterns formed
by the locations of fixations we have been able to see some
differences between the eye movements of novices and non-novices
on the Koli painting. The area that is visited most is more
concentrated and more stable in time for the non-novice group than
for the novice group. We also constructed a test to compare the
intensity surfaces of novices and non-novices. Even though the
observed difference was not statistically significant, there was
some indication that the intensities cannot be considered equal. By comparing
fixation duration distributions instead of bare means we were able
to see a clear difference between the two groups under study.
Non-novices tend to have more short fixation durations than
novices meaning that they may only glance at some areas of the
painting which novices do not tend to do. This was confirmed by the analysis based on the other paintings.

To have some idea whether the painting itself affects how novices
and non-novices look at it we compared the fixation processes of
each group on three pairs or paintings, J\"arnefelt - Monet,
Suomi - Kandinsky, and Poussin - Tammi.
Intensity surface of fixations vary of course from painting to
painting and it is not meaningful to compare them on different
paintings but it does make sense to compare the duration
distributions. An interesting observation was that in the
non-novice group the distribution of duration of fixations is more consistent from painting to painting than in the novice group.

Ideally, to be able to have a more detailed comparison of the eye
movements of novices and non-novices, we would like to be able to
describe the complete fixation process on a painting, i.e.\ where,
when and how long the gaze of a person stays in different parts of
the painting. Therefore, we have introduced a
simple reference model for the fixation process. The model is a
spatio-temporal point process model, where we used the fixation
intensity and duration distributions estimated from the data. In
order to restrict how far the gaze usually jumps we also estimated the
saccade length distribution from the data. To mimic the tendency that after having stayed on one area of the target painting for some time, the gaze jumps to another area on the painting, the model suggests a long jump with some (small) probability $p$.

We fitted the model to the non-novice and novice groups separately
for the eye movements on the J\"arnefelt painting and saw that the model fits quite well for both groups, except when measured by the scanpath length. The structure of the fixation process (described by our model) is
similar in nature for non-novices and novices. The model based simultaneous envelopes for the functional summary statistics can be used like confidence intervals to compare the two groups. However, since the envelopes overlap, we were not able to reveal any significant differences between the groups based on the model based summaries.

Our idea here was to find a rather simple model that captures the
most important features of the fixation process and the reference model is quite good for this purpose.
However, if the goal is to understand the complete dynamics of the
fixation process, the model constructed here is not good enough
since it does not capture that the
fixation process is changing in time. We noticed, for example, in Section
\ref{sec:data} that both the intensity surface and the fixation
duration distribution vary in time. Despite this we used the same
intensity surface and the same gamma distributions for the
fixation durations and saccade lengths during the whole time
period. We made some experiments by using six intensity surfaces
(for each 30 second interval) and two distributions for saccade
lengths (one for the first 30 seconds and one for the rest of the
three minute time interval) in the reference
model but these modifications did not improve the goodness-of-fit
of the  model. 

One possible way to improve the model could be to  take the order of
fixations into account. The fixation process could be regarded as a
realization of a sequential point process which can be used to
access the non-stationarity of the fixation process. The first
author of this paper is currently working on that issue (see preprint \cite{penttinen}).

To conclude, our new set of tools allowed us to make more detailed
comparisons between how novices and non-novices look at paintings
than reported in the literature. As far as we know such
comparisons of fixation intensity surfaces, duration distributions
and coverages have not been presented earlier in eye movement
studies.

\section*{Acknowledgements}
The authors are grateful to Pertti Saariluoma, Sari Kuuva, Mar\'ia
\'Alvarez Gil, Jarkko Hautala and Tuomo Kujala for providing the
data and to Antti Penttinen for helpful comments and discussions. The authors thank the two anonymous reviewers, associate editor and editor for their valuable comments that helped to significantly improve the paper.

The mobility funding provided by the University of Jyv\"askyl\"a for the first author (AKY) is highly
appreciated.

\begin{supplement}[id=suppA]
	\sname{Supplement A}
	\stitle{Statistics for fixations inside the picture.} 
	\slink[doi]{? suppmaterialA}
	\sdescription{Table includes information about the fixations, such as total number of fixations and mean fixation duration, for each subject.}
\end{supplement}

\begin{supplement}[id=suppB]
	\sname{Supplement B}
	\stitle{Statistics for saccades inside the picture.} 
	\slink[doi]{? suppmaterialB}
	\sdescription{Table includes information about the saccades, such	as mean saccade duration and mean saccade length, for each subject.}
\end{supplement}

\begin{supplement}[id=suppC]
	\sname{Supplement C}
	\stitle{The most visited areas during the six 30 second intervals for novices and non-novices.}
	\slink[doi]{? suppmaterialC} 
	\sdescription{Figures show the top 5 \% and top 1 \% intensities in 30 second intervals (0-30s), (30-60s), (60-90s), (90-120s), (120-150s), and (150-180s) for novices and non-novices.}
\end{supplement}

\begin{supplement}[id=suppD]
	\sname{Supplement D}
	\stitle{Results for the comparison of novices and non-novices for all six paintings.}
	\slink[doi]{? suppmaterialD} 
	\sdescription{Results for the intensity surface and fixation duration distribution comparisons between novices and non-novices for all six paintings used in the experiment.}
\end{supplement}

\begin{supplement}[id=suppE]
	\sname{Supplement E}
	\stitle{Results for the groupwise fixation duration distribution comparisons for three pairs of paintings}
	\slink[doi]{? suppmaterialE} 
	\sdescription{Results for the fixation duration distribution comparisons within novice and non-novice groups for the three pairs of paintings: J\"arnefelt - Monet; Kandinsky - Suomi and Poussin - Tammi.}
\end{supplement}

\bibliography{refs}
\bibliographystyle{plainnat}

\end{document}